\documentclass{amsart}
\newtheorem{lemma}{Lemma}[section]
\newtheorem{proposition}[lemma]{Proposition}
\newtheorem{corollary}[lemma]{Corollary}
\newtheorem{theorem}[lemma]{Theorem}
\theoremstyle{definition}
\newtheorem{definition}[lemma]{Definition}
\theoremstyle{definition}
\newtheorem{example}[lemma]{Example}
\newtheorem{remark}[lemma]{Remark}
\theoremstyle{definition}

\thanks{This work was supported by Science and Technology Assistance
Agency under the contract No. APVT-51-032002, grant VEGA 2/6088/26
and Center of Excellence SAS, CEPI I/2/2005}
\title {Sharp and fuzzy observables on effect algebras}
\author{Jen\v cov\'a, A., Pulmannov\'a, S., Vincekov\'a, E.}
\begin{document}
\begin{abstract} Observables on effect algebras and their fuzzy
versions obtained by means of confidence measures (Markov kernels)
are studied. It is shown that, on effect algebras with the
(E)-property, given an observable and a confidence measure, there
exists a fuzzy version of the observable. Ordering of observables
according to their fuzzy properties is introduced, and some
minimality conditions with respect to this ordering are found.
Applications of some results of classical theory of experiments are
considered.
\end{abstract}
\address{Mathematical Institute, Slovak Academy of Sciences, \v
Stef\'anikova 49, 814 73 Bratislava, Slovakia }
\email{jenca@mat.savba.sk, pulmann@mat.savba.sk,
vincek@mat.savba.sk} \keywords{effect algebra,  observable, Hilbert
space effects, PV-measure, POV-measure, sufficient Markov kernel,
smearing} \subjclass{Primary 81P10, 81P15; Secondary 62B05, 62B15}
\maketitle \markboth{Jen\v cov\'a, A. Pulmannov\'a, S., Vincekov\'a,
E.}{Sharp and fuzzy observables}
\date{}
\maketitle

\section{Introduction}
In the frame of quantum mechanics, as a proper mathematical
formulation of a physical quantity (so called observable), a
normalized positive operator valued measure is considered, instead
of the more traditional spectral measure (projection valued
measure). This approach has also provided a frame to investigate
imprecise measurements of a physical quantity. In the literature
(e.g., \cite{HLY}), the notion of a quantum mechanical fuzzy
observable has been formulated as a smearing of a sharp observable
(projection measure). In the present paper, we study smearing of
observables in a more general frame of effect algebras. In analogy
with \cite{HLY,He}, we introduce the notion of a confidence measure
(which is a Markov kernel), and we show that in a
$\sigma$-orthocomplete effect algebra with an order determining set
of $\sigma$-additive states which has the (E)-property \cite{DDD},
every confidence measure yields a smeared observable for a given
(real) observable. We can then introduce a partial order for
observables by putting $\xi \preceq \eta$ if $\eta$ is a smearing of
$\xi$; in this case we say that $\eta$ is a {\it fuzzy version} of
$\xi$ \cite{He}. If $\xi\preceq \eta$ and simultaneously, $\eta
\preceq \xi$, we will write $\xi \sim \eta$, and say that $\xi$ and
$\eta$ are {\it fuzzy equivalent}. In analogy with some recent
papers \cite{Bud}, minimal elements in this ordering are called
(postprocessing) {\it clean observables}, or {\it optimal
measurements}, \cite{He}.

As a motivation, we give the following \cite{HLY}. Let $L$ be
$\sigma$-orthocomplete effect algebra,  $(\Omega, {\mathcal A})$ a
measurable space, and $\xi: {\mathcal A}\to L$ a sharp observable on
$L$, and $m$ a $\sigma$-additive state on $L$. For every $E\in
{\mathcal A}$, $\xi(E)$ is a sharp element of $L$ (recall that $a\in
L$ is sharp if $0$ is the unique common lower bound of $a$ and its
orthosupplement $a'$), and $m(\xi(E))=\int \delta_{\omega}(E)
m(\xi(dx))$, where $\delta_{\omega}(E)=1$ if $\omega\in E$, and
$\delta_{\omega}(E)=0$ if $\omega\notin E$. Since realistic
measurements always have some imprecision, one may think that the
points of $\Omega$ are to be replaced by probability distributions.
If we replace the Dirac function $\delta_{\omega}$ by a probability
distribution $\nu_{\omega}:{\mathcal A}\to [0,1]$ in every point
$\omega$, we obtain $\int_{\Omega}\nu_{\omega}(E)m(\xi(d\omega))$,
which is a smearing of $m(\xi(E))$. Under some appropriate
additional assumptions on $L$ (which are satisfied in the case of
the effect algebra ${\mathcal E}(H)$ of the Hilbert space effects),
there is a smeared observable $\eta$ of $\xi$, such that
$m(\eta(E))=\int_{\Omega}\nu_{\omega}(E)m(\xi(d\omega))$ for every
$E\in {\mathcal A}$ and every $\sigma$-additive state $m$.

\section{Effect algebras}

An {\it effect algebra} \cite{FB} (see \cite{GG} and \cite{KCh} for
alternative definitions) is a set $L$ with two distinguished
elements $0,1$ and with a partial binary operation $\oplus :L\times
L \to L$ such that for all $a,b,c \in L$ we have
\begin{enumerate}
\item[{\rm(EAi)}] if $a\oplus b$ exists in $L$ then $b\oplus a$
exists in $L$ and $a\oplus b=b\oplus a$ (commutativity);
\item[{\rm(EAii)}] if $b\oplus c$ exists in $L$ and $a\oplus(b\oplus
c)$ exists in $L$ then $a\oplus b$ exists in $L$ and $(a\oplus
b)\oplus c$ exists in $L$, and $(a\oplus b)\oplus c=a\oplus(b\oplus
c)$ (associativity);
\item[{\rm(EAiii)}] for every $a\in L$ there is a unique $b\in L$
such that $a\oplus b=1$ (orthosupplementation);
\item[{\rm(EAiv)}] if $1\oplus a$ is defined, then $a=0$ (zero-one
law).
\end{enumerate}

As usual, we shall write $L=(L;\oplus,0,1)$ for effect algebras. If
the assumptions of (EAii) are satisfied, we write $a\oplus b\oplus
c$ for the element $(a\oplus b)\oplus c=a\oplus(b\oplus c)$ in $L$.

Let $a,b$ be elements of an effect algebra $L$. We say that (i) $a$
is {\it orthogonal to $b$} and write $a\perp b$ iff $a\oplus b$ is
defined in $L$; (ii) $a$ is {\it less than or equal to} $b$ and
write $a\leq b$ iff there exists an element $c$ in $L$ such that
$a\perp c$ and $a\oplus c=b$ (in this case we also write $b\geq a$);
$b$ is the {\it orthosupplement} of $a$ and write $b=a'$ iff $b$ is
the (unique) element in $L$ such that $b\perp a$ and $a\oplus b=1$.
If $a\leq b$, then the element $c$ such that $a\oplus c=b$ is
uniquely defined, and we write $c=b\ominus a$. In particular, for
every $a\in L$, $a'=1\ominus a$, $a\perp b$ iff $b\leq a'$, and
$(a\oplus b)'=a'\ominus b$.

For a finite sequence $a_1,a_2,\ldots,a_n$, $n\geq 3$, we define
recursively
\begin{equation}\label{eq:1}
a_1\oplus \cdots \oplus a_n:=(a_1\oplus \cdots \oplus a_{n-1})\oplus
a_n,
\end{equation}
supposing that $(a_1\oplus \cdots \oplus a_{n-1})$ and $(a_1\oplus
\cdots \oplus a_{n-1})\oplus a_n$ exist in $L$. Due to associativity
of $\oplus$, the element (\ref{eq:1}) is correctly defined. Define
$a_1\oplus \cdots \oplus a_n=a_1$ if $n=1$, and $a_1\oplus \cdots
\oplus a_n=0$ if $n=0$. Then, due to commutativity and associativity
of $\oplus$,  for any permutation $(i_1,i_2,\ldots,i_n)$ of
$(1,2,\ldots,n)$ and any $0\leq k\leq n$ we have
\begin{equation}\label{eq:2}
a_1\oplus \cdots \oplus a_n=a_{i_1}\oplus \cdots \oplus a_{i_n},
\end{equation}
\begin{equation}\label{eq:3}
a_1\oplus \cdots \oplus a_n=(a_1\oplus \cdots \oplus
a_k)\oplus(a_{k+1}\oplus \cdots \oplus a_n).
\end{equation}

We say that a finite sequence $F=\{ a_1,\ldots, a_n\}$ is {\it
orthogonal} if $a_1\oplus \cdots \oplus a_n$ exists in $L$, and we
say that {\it $F$ has the $\oplus$-sum $\bigoplus F$}, which is
defined by
\begin{equation}\label{eq:4}
\bigoplus F=a_1\oplus \cdots \oplus a_n.
\end{equation}

An arbitrary system $G=\{ a_i\}_{i\in I}$ of (not necessarily
different) elements of $L$ is said to be {\it orthogonal} if for any
finite subset $J$ of $I$, the system $\{ a_i\}_{i\in J}$ is
orthogonal. An orthogonal system $G=\{ a_i\}_{i\in I}$ has an {\it
$\oplus$-sum} in $L$, if in $L$ there exists the join
\begin{equation}\label{eq:5}
\bigoplus_{i\in I}a_i:=\bigvee_J \bigoplus_{i\in J} a_i,
\end{equation}
where $J$ runs over all finite subsets of $I$. In this case, we also
write $\bigoplus G:=\bigvee_J \bigoplus_{i\in I}a_i$.

Evidently, if $G=\{ a_1,\ldots,a_n\}$ is orthogonal, then the
$\oplus$-sums defined by (\ref{eq:4}) and (\ref{eq:5}) coincide.

Let $G=\{ a_i\}_{i\in I}$ and $a_i=a$ for all $i\in I$. The greatest
$n$ such that $\bigoplus_{i\leq n} a_i$ exists, is called the {\it
isotropic index} of $a$. If $\bigoplus_{i\leq n} a_i$ exists for all
$n\in {\mathbb N}$, we say that the isotropic index of $a$ is
infinite. If $\bigoplus G$ exists and $I$ is infinite, then $a=0$.
Indeed, let $a_0=\bigoplus G$, then $a_0=a_j\oplus \bigoplus_{i\in
I\setminus \{ j\}}a_i=a\oplus a_0$, which gives $a=0$. Notice that
if $G$ is only orthogonal, then $a$ is not necessarily $0$.

We say that an effect algebra $L$ is {\it $\sigma$-orthocomplete}
({\it orthocomplete}) if $\bigoplus_{i\in I}a_i$ exits for any
countable (arbitrary) orthogonal system $\{ a_i:i\in I\}$ of
elements of $L$. We recall that an effect algebra is
$\sigma$-orthocomplete iff for every nondecreasing sequence $\{
a_i\}_{i\in {\mathbb N}}$ there is a supremum $a=\bigvee_{i\in
{\mathbb N}} a_i$.

A mapping $s:L\to [0,1]$ from $L$ to unit interval $[0,1]$ of real
numbers is a {\it state} on $L$ if (i) $s(1)=1$, (ii) $s(a\oplus
b)=s(a)+s(b)$ whenever $a\oplus b$ exists in $L$. It is clear that
$s(0)=0$, and $s(a)\leq s(b)$ whenever $a\leq b$, $a,b\in L$. A
state $s:L\to [0,1]$ is said to be {\it $\sigma$-additive}, or {\it
completely additive} if the equality
\begin{equation}\label{eq:6}
s(\bigoplus_{i\in I}a_i)=\sum_{i\in I} s(a_i),
\end{equation}
holds for any countable, or arbitrary index set $I$, respectively,
such that $\bigoplus_{i\in I}a_i$ exists in $L$.

A non-void system ${\mathcal S}$ of states on $L$ is said to be {\it
order determining}, if for $a,b\in L$, $a\leq b$ iff $s(a)\leq s(b)$
for all $s\in {\mathcal S}$. We denote by $Conv({\mathcal S})$ and
$Conv_{\sigma}({\mathcal S})$ the convex and $\sigma$-convex hull of
${\mathcal S}$, respectively. Clearly, elements of $Conv({\mathcal
S})$ and $Conv_{\sigma}({\mathcal S})$ are states on $L$, and
moreover, $\mathcal S$ is order determining iff $Conv({\mathcal S})$
is order determining, or, equivalently, iff $Conv_{\sigma}({\mathcal
S})$ is order determining.

Let $L$ and $P$ be effect algebras, a mapping $\phi:L\to P$ is a
{\it morphism} if (i) $m(1_L)=1_P$, where $1_L$ and $1_P$ are the
unit elements in $L$ and $P$, respectively, and (ii) $a\perp b$
implies $\phi(a)\perp \phi(b)$, and $\phi(a\oplus b)=\phi(a)\oplus
\phi(b)$. A morphism $\phi$ is called a {\it $\sigma$-morphism}
({\it complete morphism}) if it preserves all existing countable
(arbitrary) $\oplus$-sums. A bijective morphism such that $a\perp b$
iff $\phi(a)\perp \phi(b)$, is an {\it isomorphism}. A
$\sigma$-isomorphism, resp. complete isomorphism, is defined in an
obvious way.

A subset $P$ of  an effect algebra $L$ is a {\it sub-effect
algebra}, if (i) $0\in P$, $1\in P$; (ii) $a,b\in P$, $a\perp b$
implies $a\oplus b\in P$, (iii) $a\in P$ implies $a'\in P$.

We recall that an effect algebra is:
\begin{enumerate}
\item[-] an {\it orthoalgebra} iff
$a\perp a$ implies $a=0$;
\item[-] an {\it orthomodular poset} iff $a\perp b$ implies $a\oplus
b=a\vee b$;
\item[-] an {\it orthomodular lattice} iff it is a lattice ordered
orthomodular poset;
\item[-] an {\it MV-effect algebra} iff it is lattice ordered and the
equalities $(a\vee b)\ominus a=b\ominus(a\wedge b)$ are satisfied.
We recall that MV-effect algebras coincide with {\it MV-algebras}
introduced by Chang \cite{Ch} as algebraic bases for many-valued
logic.
\end{enumerate}

Two of the most important prototypes of effect algebras are the
following examples.

\begin{example}\label{ex:[0,1]} Consider the closed interval $[0,1]$
of reals ordered by the natural way. For two numbers $a,b\in [0,1]$
define $a\oplus b$ iff $a+b\leq 1$ and put then $a\oplus b=a+b$.
Then $[0,1]$ is an orthocomplete effect algebra, and the effect
algebra order coincides with the natural order of reals. With
respect to this order, $[0,1]$ is a totally ordered, distributive
lattice. We recall that $\{ a_t\}$ is orthogonal iff $\sum_ta_t\leq
1$, and $\bigoplus_t a_t=\sum_t a_t$. There is only one state on
$[0,1]$, namely the isomorphism $s_0(a)=a$. Clearly, $s_0$ is
completely additive and the one-point set $\{ s_0\}$ is order
determining.

We recall that $[0,1]$ is also a prototype of MV-algebras.
\end{example}

\begin{example}\label{ex:effect}
 The set ${\mathcal E}(H)$ of all self-adjoint
operators $A$ on a Hilbert space $H$ such that $0\leq A\leq I$,
where $0$ is the zero and $I$ the identity mapping, ordered by the
usual order of self-adjoint operators, namely $A\leq B$ iff $\langle
Ax,x\rangle\leq \langle Bx,x\rangle$ for all $x\in H$. We define, on
${\mathcal E}(H)$, $A\perp B$ iff $A+B\leq I$, and then put $A\oplus
B=A+B$. Then $({\mathcal E}(H);\oplus,0,I)$ becomes an effect
algebra, in which the algebraic order coincides with the usual order
that we started with. A system $\{ A_t\}_t$ of elements from
${\mathcal E}(H)$ is orthogonal if $\sum_t A_t\leq I$, where the
summation is in the weak, or equivalently in the strong operator
topology, and then $\bigoplus_t A_t=\sum_t A_t$. The system
$({\mathcal E}(H);\oplus,0,I)$ is an orthocomplete effect algebra
which is not a lattice \cite{LM, Gud}.

Denote by ${\mathcal P}(H)$ the set of all orthogonal projections on
$H$. Then ${\mathcal P}(H)$ is a sub-effect algebra of ${\mathcal
E}(H)$, which is a complete orthomodular lattice.

We recall that ${\mathcal E}(H)$, as well as ${\mathcal P}(H)$, play
an important role in the foundations of quantum mechanics and the
theory of quantum measurements \cite{BLM}.
\end{example}

\section{Observables on effect algebras}

Let $L$ be a $\sigma$-orthocomplete effect algebra, and
$(\Omega,{\mathcal A})$ a measurable space. By an  {\it
$(\Omega,{\mathcal A})$- observable} on $L$ we mean a mapping
$\xi:{\mathcal A} \to L$ such that
\begin{enumerate}
\item[(i)] $\xi(\Omega)=1$;
\item[(ii)] the system $\{ \xi(E_i)\}_{i\in {\mathbb N}}$ is orthogonal and $\xi(\bigcup_{i=1}^{\infty}
E_i)=\bigoplus_{i=1}^{\infty}\xi(E_i)$ whenever  $E_i\cap
E_j=\emptyset$, $i\neq j$, and $E_i\in {\mathcal A}$ for $i\geq 1$.
\end{enumerate}

If $(\Omega,{\mathcal A})\subseteq ({\mathbb R}, {\mathcal
B}({\mathbb R}))$, then an  observable $\xi:{\mathcal A} \to L$ is
said to be a {\it real} observable.

Let $(\Omega_1,{\mathcal A}_1)$ be another measurable space, and let
$f:\Omega \to \Omega_1$ be a measurable function such that
$f^{-1}(A)\in {\mathcal A}$ whenever $A\in {\mathcal A}_1$. If
$\xi:{\mathcal A}\to L$ is an observable, then $f\circ \xi:A\mapsto
\xi(f^{-1}(A))$, $A\in {\mathcal A}_1$ is a $(\Omega_1,{\mathcal
A}_1)$-observable on $L$. It is called the {\it $f$-function of
$\xi$}. In particular, if $\xi$ is a real observable on $L$ and
$f:{\mathbb R}\to {\mathbb R}$ is a Borel measurable function, then
$f\circ \xi$ is also a real observable on $L$.

If $\xi$ is a $(\Omega,{\mathcal A})$-observable on $L$, and $s$ is
a $\sigma$-additive state on $L$, then $s_{\xi}:=s\circ
\xi:{\mathcal A}\to [0,1]$ is a probability measure on
$(\Omega,{\mathcal A})$. If $\xi$ is a real observable, we denote by
\begin{equation}\label{eq:7}
s(\xi):=\int_{\mathbb R}ts_{\xi}(dt)
\end{equation}
the {\it mean value} of $\xi$ in $s$ whenever the right-hand side of
the above equation exists and is finite.

More generally, if $\xi:(X,{\mathcal A})\to L$ is an observable,
then for any Borel measurable function $f:X\to {\mathbb R}$,
$f(\xi)$ is a real observable, and
\begin{eqnarray*}
s(f(\xi))&=&\int_{\mathbb R}u s(f(\xi(du))\\
&=&\int_{\mathbb R}u s(\xi(f^{-1}(du))) =\int_Xf(t)s_{\xi}(dt)),
\end{eqnarray*}
using the integral transformation theorem.

The {\it spectrum} of a real observable  $\xi$ is the smallest
closed subset $C$ of ${\mathbb R}$ such that $\xi(C)=1$.

For an $(\Omega,{\mathcal A})$-observable $\xi$ on $L$, let
${\mathcal R}(\xi):=\{ \xi(A):A\in {\mathcal A}\}$ denote the range
of $\xi$. Recall that an element $a\in L$ is called {\it sharp} if
$a\wedge a'=0$, that is, $0$ is the only common lower bound of $a$
and $a'$. Clearly, $0,1$ are sharp, and $a$ is sharp iff $a'$ is
sharp. We will say that and observable $\xi$ is {\it sharp} if its
range consists of sharp elements.

Let us consider the following examples.

\begin{example}\label{ex:pov}
 Let $H$ be a Hilbert space, and ${\mathcal E}(H)$
be the effect algebra of Example \ref{ex:effect}. Here the sharp
elements coincide with projections. Indeed, if $P$ is a projection,
then $P=P^2$ implies $P\wedge (I-P)=P(I-P)=0$, and conversely, for
any $A\in {\mathcal E}(H)$, $0\leq A\leq I$ implies that
$A^{\frac{1}{2}}AA^{\frac{1}{2}}\leq
A^{\frac{1}{2}}IA^{\frac{1}{2}}$, which yields $0\leq A^2\leq A$.
Then $0\leq A-A^2\leq A$, and $(I-A)-(A-A^2)=(I-A)^2\geq 0$ yields
$I-A\geq A-A^2$, hence $A-A^2$ is a common lower bound of $A$ and
$I-A$. Hence, $A$ is sharp iff $A=A^2$, equivalently, iff $A$ is a
projection. Sharp observables on ${\mathcal E}(H)$ are then exactly
those whose ranges are in ${\mathcal P}(H)$. These observables are
called {\it projection valued observables} (PV-observables, in
short), while general observables are called {\it positive operator
valued observables} (POV-observables, in short). Owing to spectral
theorem, real (bounded) PV-observables are in one-to-one
correspondence with (bounded) self-adjoint operators.
\end{example}

 \begin{example}\label{ex:tribe} Let $X$ be a nonempty set. A {\it tribe} over $X$
is a collection of functions ${\mathcal T}\subseteq [0,1]^X$such
that the zero function ${\underline 0}(x)=0$ is in ${\mathcal T}$
and the following is satisfied:
\begin{enumerate}
\item[{\rm(T1)}] $f\in {\mathcal T}\ \implies \, 1-f\in {\mathcal T}$;
\item[{\rm(T2)}] $f,g\in {\mathcal T}\, \implies \, f\dot{+} g:=\min(f+g,1)\in
{\mathcal T}$;
\item[{\rm(T3)}] $f_n\in {\mathcal T}, n\in {\mathbb N}$ and $ f_n\nearrow f$
(pointwise) $\implies \, f\in {\mathcal T}$.
\end{enumerate}

Elements of ${\mathcal T}$ are called {\it fuzzy sets} or {\it fuzzy
events}. Sharp elements in ${\mathcal T}$ coincide with the
characteristic functions contained in ${\mathcal T}$. We put
${\mathcal B}({\mathcal T}):=\{ B\subseteq X:\chi_B\in {\mathcal
T}\}$, where $\chi_B$ is the characteristic function of the set $B$.
Then ${\mathcal B}({\mathcal T})$ is a $\sigma$-algebra of sets,
which is isomorphic with the system of all sharp elements of
${\mathcal T}$. The restriction of any $\sigma$-additive state on
$\mathcal T$ to ${\mathcal B}({\mathcal T})$ is a probability
measure. Due to Butnariu and Klement theorem \cite{BK}, every
element in ${\mathcal T}$ is a measurable function with respect to
${\mathcal B}({\mathcal T})$. Moreover, every $\sigma$-additive
state $m$ on ${\mathcal T}$ has an integral representation
\begin{equation}\label{eq:8}
m(f)=\int_X fdP,
\end{equation}
where $P$ is the restriction of $m$ to ${\mathcal B}({\mathcal T})$,
i.e., $P(A)=m(\chi_A)$.

Let $\xi$ be an $(\Omega,{\mathcal A})$-observable on ${\mathcal
T}$. Define $\nu :X\times {\mathcal A}\to [0,1]$,
$\nu(x,A)=\xi(A)(x)$, where $\xi(A)\in {\mathcal T}$. The mapping
$\nu$ has the following properties:
\begin{enumerate}
\item[{\rm(c1)}] for any fixed $x\in X$, $\nu(x,.)$ is a probability measure
on ${\mathcal A}$;
\item[{\rm(c2)}] for any fixed $A\in {\mathcal A}$, $\nu(.,A)$ belongs to
${\mathcal T}$.
\end{enumerate}
Conversely, every mapping $\nu :X\times {\mathcal A}\to [0,1]$ with
properties (c1), (c2) gives rise to an observable on ${\mathcal T}$
given by $\xi(A)=\nu(.,A)$.

Clearly, an observable is sharp if its range consists of
characteristic functions from ${\mathcal T}$. In fact, if $\pi$ is a
sharp $(\Omega,{\mathcal A})$-observable on ${\mathcal T}$, then
$\pi$ is a $\sigma$-homomorphism $\pi :{\mathcal A}\to {\mathcal
B}({\mathcal T})$, therefore there is an $({\mathcal A},{\mathcal
B}({\mathcal T}))$-measurable function $g:X\to \Omega$ such that
$\pi(A)=g^{-1}(A)$, $A\in {\mathcal A}$, and
$\nu(x,A)=\chi_{g^{-1}(A)}(x)$ (\cite{Var}).
\end{example}

\begin{example}\label{ex:mv} Recall that an MV-algebra can be defined as a system
$(M,\dot{+},*,0,1)$ consisting of a nonempty set $M$, two constants
$0$ and $1$, a unary operation $^*$ and a binary operation $\dot{+}$
satisfying the following axioms:
\begin{enumerate}
\item[{\rm(MV1)}] $a\dot{+} b=b\dot{+} a$;
\item[{\rm(MV2)}] $a\dot{+}(b\dot{+}c)=(a\dot{+}b)\dot{+}c$;
\item[{\rm(MV3)}] $a\dot{+} a^*=1$;
\item[{\rm(MV4)}] $a\dot{+} 0=a$;
\item[{\rm(MV5)}] ${a^*}^*=a$;
\item[{\rm(MV6)}] $0^*=1$;
\item[{\rm(MV7)}] $a\dot{+} 1=1$;
\item[{\rm(MV8)}] $(a^*\dot{+} b)^*\dot{+}b=(a\dot{+}b^*)^*\dot{+}
a$.
\end{enumerate}

The above axioms are equivalent with the original axioms introduced
by Chang in \cite{Ch} (see \cite{CDM}). A partial order can be
introduced on $M$ by putting $a\leq b$ iff $a^*\dot{+} b=1$. With
respect to this ordering, $M$ becomes a distributive lattice, where
$a\vee b=(a^*\dot{+} b)^*\dot{+} b$, $a\wedge b=(a^*\vee b^*)^*$. By
putting $a\oplus b=a\dot{+} b$ iff $a\leq b^*$, we obtain an effect
algebra $(M;\oplus,0,1)$, where $a^*$ is the orthosupplement of $a$
for all $a\in M$. Conversely, an effect algebra $(L;\oplus,0,1)$ can
be organized into an MV-algebra (i.e., it is an MV-effect algebra)
iff $L$ is a lattice, and for any $a,b\in L$, the equality $(a\vee
b)\ominus a=b\ominus(a\wedge b)$ holds. The total operation
$\dot{+}$ is defined by $a\dot{+}b=(a\oplus (a'\wedge b))$, and
$a^*=a'$ \cite{ChK}. An MV-effect algebra $M$ is
$\sigma$-orthocomplete ($\sigma$-MV algebra), or orthocomplete
(complete MV algebra) iff $M$ is a $\sigma$-lattice, or a complete
lattice, respectively. Sharp elements in an MV-algebra $M$ coincide
with the idempotents in $M$, that is, $a\wedge a^*=0$ iff $a\dot{+}
a=a$. The set ${\mathcal B}(M)$ of sharp elements of $M$ forms a
Boolean subalgebra of $M$. If $M$ is $\sigma$-complete, then
${\mathcal B}(M)$ is a Boolean $\sigma$-algebra \cite{DvPu}.

Every tribe is a $\sigma$-MV algebra with $f\dot{+} g=\min(f+g,1)$,
$f^*=1-f$, and where the lattice operations $\vee, \wedge$ coincide
with pointwise supremum and infimum, respectively, of $[0,1]$-valued
functions on $X$.

By the Loomis-Sikorski theorem for $\sigma$-MV algebras \cite{Mu},
\cite{Dv}, \cite{BW}, to every $\sigma$-MV algebra there is a triple
$(X,{\mathcal T},h)$ consisting of a tribe $\mathcal T$ of fuzzy
sets on a nonvoid set $X$ and a surjective $\sigma$-homomorphism (of
$\sigma$-MV-algebras) $h:{\mathcal T}\to M$, such that the
restriction of $h$ to ${\mathcal B}({\mathcal T})$ maps the latter
set onto ${\mathcal B}(M)$.

Let $M$ be a $\sigma$-MV-effect algebra, and let $(X,{\mathcal
T},h)$ be its representation by the Loomis-Sikorski theorem.  Let
$\xi$ be an $(\Omega,{\mathcal A})$ observable on $M$. For every
$A\in {\mathcal A}$, there is an $f_A\in {\mathcal T}$ such that
$h(f_A)=\xi(A)$, where $f_A$ is unique up to $h$-null sets.  Define
$\nu :X\times {\mathcal A}\to [0,1]$ by putting $\nu(x,A)=f_A(x)$.
Clearly, for a fixed $A\in {\mathcal A}$, $\nu_A\in {\mathcal T}$.
Moreover, $\xi(A)=h(\nu_A)$. Let $\{ E_i\}_i$ be a disjoint sequence
of elements of ${\mathcal A}$, and put $E=\bigcup_i E_i$. Then
$\xi(E)=\bigoplus_i \xi(E_i)$. Choose functions $f, f_i,
i=1,2,\ldots$ in ${\mathcal T}$ such that $h(f)=\xi(E)$,
$h(f_i)=\xi(E_i)$, $i=1,2,\ldots$. Then we have $h(f)=\bigoplus
h(f_i)=h(\min(\sum_{i=1}^{\infty}f_i,1))$. Consider an orthogonal
sequence $g_i, i=1,2,\ldots$, where $g_1=f_1$, and for $i\geq 1$,
$g_i=f_i\wedge(g_1+\ldots+g_{i-1})^*$. We have $h(f_1)=h(g_1)$, and
assume that $h(g_i)=h(f_i)$ for $i<k$. Then
$h(g_k)=h(f_k\wedge(g_1+\ldots g_{k-1})^*)=h(f_k)\wedge h(g_1+\ldots
+g_{k-1})'=h(f_k)\wedge (h(f_1)\oplus \ldots \oplus
h(f_{k-1}))'=h(f_k)$. We proved, by induction, that $h(f_i)=h(g_i),
i=1,2,\ldots$, which entails that $h(\{ x:f_i(x)\neq g_i(x)\})=0$,
$i=1,2,\ldots$ (we identify sets with their characteristic
functions). Clearly, $\sum_{i=1}^{\infty}f_i>1$ iff $f_i\neq g_i$
for at least one $i$, therefore $\{ x:
\sum_{i=1}^{\infty}f_i>1\}=\bigcup_{i=1}^{\infty} \{ x:f_i\neq
g_i\}\in \ker(h)$, and this entails that $h(\{ x: f(x)\neq
\sum_{i=1}^{\infty}f_i(x)\})=0$. This shows that
$\nu(x,E)=\sum_{i=1}^{\infty}\nu(x,E_i)$ for all $x$ up to an
$h$-null set.
\end{example}

\section{Smearing of observables}

\subsection{Markov kernels}

Let $L$ be a $\sigma$-orthocomplete effect algebra with a system
 ${\mathcal S}$ of $\sigma$-additive states, and let
$(X,{\mathcal F})$ and $(Y,{\mathcal G})$ be measurable spaces. Let
$\xi$ be an $(X,{\mathcal F})$-observable on $L$. Consider a mapping
$\nu:X\times {\mathcal G}\to [0,1]$ with the following properties:
\begin{enumerate}
\item[(i)] for any fixed $x\in X$, $\nu_x(.):=\nu(x,.):{\mathcal G}\to [0,1]$ is
a probability measure;
\item[(ii)] for any fixed $G\in {\mathcal G}$, the mapping $x\mapsto
\nu_G(x):= \nu(x,G)$ is ${\mathcal F}$-measurable.
\end{enumerate}
That is, $\nu$ is a {\it Markov kernel} (we note that in analogy
with \cite{HLY}, $\nu$ may be called also a {\it confidence
measure}). Let $m\in {\mathcal S}$. The integral
$$
\int_X\nu_G(x)m(\xi(dx))
$$
converges by the dominating convergence theorem. If there is an
observable $\eta :(Y,{\mathcal G})\to L$ such that
\begin{equation}\label{eq:9}
m(\eta(G))=\int_X\nu_G(x)m(\xi(dx))
\end{equation}
for every $m\in {\mathcal S}$, the we will call $\eta$ a {\it fuzzy
version} of $\xi$, or a {\it smearing of $\xi$ in the states $m\in
{\mathcal S}$}. If moreover the system ${\mathcal S}$ is order
determining, then the equations (\ref{eq:9}) uniquely determine
$\eta$, and  we call $\eta$ simply a fuzzy version (smearing) of
$\xi$. In this case, we will write $\xi \preceq \eta$. If equation
(\ref{eq:9}) holds for every
 $m\in {\mathcal S}$, we will write symbolically
\begin{equation}\label{eq:9a}
\eta(G)=\int_X\nu_G(x)\xi(dx).
\end{equation}

The relation $\preceq$ is reflexive, since the mapping $(x,G)\mapsto
\delta_x(G)=\chi_G(x)$ is a Markov kernel, and $\xi(G)=\int
\chi_G(x)\xi(dx)$. It is also transitive. Indeed, let $\xi \preceq
\eta$ and $\eta \preceq \zeta$, where $\xi :(X, {\mathcal F})\to L$,
$\eta :(Y,{\mathcal G})\to L$, $\zeta :(Z,{\mathcal H})\to L$,
$\eta(G)=\int_X \nu_1(x,G)\xi(dx)$, $\zeta(H)=\int_Y
\nu_2(y,H)\eta(dy)=\int_Y \nu_2(y,H)\int_X \nu_1(x,dy)\xi(dx)$. It
is well known that
\begin{equation}\label{eq:9b}
\nu_3(x,H):= \int_Y \nu_2(y,H)\nu_1(x,dy)
\end{equation}
is a Markov kernel (see \cite{He} for a detailed proof). Therefore,
$\preceq$ is a preorder, and it can be made a partial order in the
usual way. If $\xi \preceq \eta$, and $\eta \preceq \xi$, we will
write $\xi \sim \eta$, and we will say that $\xi$ and $\eta$ are
{\it fuzzy equivalent}. Obviously, $\xi$ is a minimal element with
respect to $\preceq$ if  $\eta \preceq \xi$ implies $\eta \sim \xi$.
Minimal observables are called {\it clean} in accordance with
\cite{Bud}. We note that in \cite{He}, the relation $\preceq$ is
defined in the opposite direction, and maximal elements are called
optimal measurements.

\subsection{Weak Markov kernels}

Let $(\Omega, {\mathcal A})$ be a measurable space. Then
$M_1^+(\Omega, {\mathcal A}))$ will denote the set of all
probability measures on $(\Omega, {\mathcal A})$.

The notion of a Markov kernel can be weakened as follows. Let
$(\Omega, {\mathcal A})$ and $(\Omega_1, {\mathcal A}_1)$ be
measurable spaces. Let ${\mathcal P}\subseteq M_1^+(\Omega,
{\mathcal A}))$, and let $\nu: \Omega \times {\mathcal A}_1 \to
{\mathbb R}$. We will say that $\nu$ is a {\it weak Markov kernel
with respect to ${\mathcal P}$} if
\begin{enumerate}
\item[(i)] $\omega \mapsto \nu(\omega, B)$ is ${\mathcal
A}$-measurable for all $B\in {\mathcal A}_1$;
\item[(ii)] for every $B\in {\mathcal A}_1$, $0\leq \nu(\omega,
B)\leq 1,\,  {\mathcal P}$-a.e.;
\item[(iii)] $\nu(\omega, \Omega_1)=1,\,  {\mathcal P}$-a.e. and
$\nu(\omega,\emptyset)=0,\, {\mathcal P}$-a.e. .
\item[(iv)] if $\{ B_n\}$ is a sequence in ${\mathcal A}_1$ such
that $B_n\cap B_m=\emptyset$ for $m\neq n$, then
$$
\nu(\omega, \bigcup_n B_n)=\sum_n \nu(\omega, B_n),\, {\mathcal
P}-a.e. .
$$
\end{enumerate}

Note that a weak Markov kernel with respect to the whole
$M_1^+(\Omega,{\mathcal A})$ is in fact a Markov kernel.

It is easy to see that if $\nu$ is a weak Markov kernel with respect
to ${\mathcal P}$, then
\begin{equation}\label{eq:mk}
\nu(P)(B):=\int_{\Omega} \nu(\omega, B)P(d\omega), \, B\in {\mathcal
A}_1
\end{equation}
is a probability measure on ${\mathcal A}_1$ for all probability
measures $P\in {\mathcal P}$.

Let $L$ be a $\sigma$-orthocomplete effect algebra with an order
determining system of $\sigma$-additive states ${\mathcal S}$, and
let $(X,{\mathcal F})$ and $(Y,{\mathcal G})$ be measurable spaces.
Let $\xi$ be a $(X,{\mathcal F})$-observable on $L$. If $\nu:X\times
{\mathcal G}\to {\mathbb R}$ is a weak Markov kernel with respect to
${\mathcal P}= \{ m\circ \xi: m\in {\mathcal S}\}$, then
$$
\nu(m\circ \xi)(B)=\int_X\nu(\omega,B)m\circ \xi(dx)
$$
is a probability measure on $(Y,{\mathcal G})$, and if there is an
observable $\eta$ on $L$ such that
$$
\nu(m\circ \xi)(B)=m(\eta(B)),
$$
for all $B\in {\mathcal G}$ and all $m\in {\mathcal S}$, then we
will also call $\eta$ a {\it fuzzy version} (or a {\it smearing}) of
$\xi$  (in the states $m\in {\mathcal S}$, if the latter set is not
order determining). If ${\mathcal S}$ is order determining, we also
write $\xi\preceq \eta$.

\begin{remark}\label{re:reg} We note that a weak Markov kernel
$\nu:X\times {\mathcal G}\to[0,1]$ (with respect to one probability
measure $P$) is called a {\it random measure} in the literature. If
${\mathcal G}$ is the Borel $\sigma$-algebra of subsets of a
complete separable metric space $Y$, then there exists a regular
version $\nu^*$ of $\nu$, such that $\nu^*$ is a Markov kernel, and
\begin{equation}\label{eq:reg} \forall G\in {\mathcal G},  \nu(x,G)=\nu^*(x,G),\,\, a.e. P
\end{equation}
(see, e.g. \cite[VI.1. 21.]{St}). For a more general version, see
Theorem \ref{th:stand}.

Let $L$ be a $\sigma$-orthocomplete effect algebra with an order
determining set of $\sigma$ additive states ${\mathcal S}$, $\xi$ be
an $(X,{\mathcal F})$-observable  on $L$, and  $(Y,{\mathcal G})$ be
a complete metric space with the Borel $\sigma$-algebra. Let
$\nu:X\times {\mathcal G}\to [0,1]$ be a weak Markov kernel with
respect to ${\mathcal P}=\{ m\circ \xi: m\in {\mathcal S}\}$. Then
for every $m\in {\mathcal S}$ there exists a Markov kernel
$\nu^*_m$.

If, in addition, there is a faithful state\footnote{We recall that a
state $m_0$ on an effect algebra $L$ is {\it faithful} if $m_0(a)=0
\implies a=0$. Clearly, for every state $m$ on $L$, $m_0(a)=0\
\implies \ m(a)=0$ ($a\in L$), whence for every observable $\xi$ it
holds $m\circ \xi \ll m_0\circ \xi$. For example, if $H$ is a
complex, separable Hilbert space, then there exists a faithful state
$m_0$ on ${\mathcal E}(H)$.} $m_0$ on $L$, then the regular version
$\nu^*_{m_0}$ is the regular version of $\nu$  for all $m\in
{\mathcal S}$. This will follow from Theorem \ref{th:stand}.
\end{remark}

\begin{example}\label{ex:mk}

1. We can see that in Examples \ref{ex:tribe} and \ref{ex:mv} Markov
kernels  are closely related to observables. Namely, in Example
\ref{ex:tribe} observables coincide with certain Markov kernels.

In Example \ref{ex:mv}, we have the following situation. Let $(X,
{\mathcal T}, h)$ be the Loomis-Sikorski representation of a
$\sigma$-MV-algebra $M$. Let $\xi :(\Omega, {\mathcal A})\to M$ be
an observable. For every $A\in {\mathcal A}$, choose $f_A\in
{\mathcal T}$ such that $\xi(A)=h(f_A)$ (for definiteness, we may
choose the (unique) continuous function in the corresponding class),
and define $\nu(x,A)=f_A(x)$, then $\nu:X\times {\mathcal A}\to
[0,1]$ is a weak Markov kernel with respect to the family ${\mathcal
P}=\{ m\circ h\}$, $m$  a $\sigma$-additive state on $M$, of
probability measures on ${\mathcal B}({\mathcal T})$. Owing to
Butnariu-Klement theorem we have
$$
m(\xi(A))=m(h(f_A))=\int_Xf_A(x)P(dx),
$$
where $P=m\circ h/{\mathcal B}({\mathcal T})=m\circ(h/{\mathcal
B}({\mathcal T}))$. The restriction $h/{\mathcal B}({\mathcal
T}):{\mathcal B}({\mathcal T})\to {\mathcal B}(M)$ can be considered
as a sharp observable on $M$, and any other observable may be
considered as a smearing of it in all $\sigma$-additive states on
$M$.

2. Let $\eta, \xi$ be real observables on an effect algebra $L$ such
that $\eta=f\circ \xi$ for some Borel function $f:{\mathbb R}\to
{\mathbb R}$. Then for every $\sigma$-additive state $m$ on $L$,
$m(\eta(E))=m(\xi(f^{-1}(E))=\int
\chi_{f^{-1}(E)}(\lambda)m(\xi(d\lambda))$, $E\in {\mathcal
B}({\mathbb R})$. Put $\nu(\lambda,E)=\chi_{f^{-1}(E)}(\lambda)$. It
is easy to see that $\nu$ is a Markov kernel, and $\eta$ is a
smearing of $\xi$. However, if $\xi$ is sharp, then $\eta$ is sharp,
too. Hence a smearing of a sharp observable may be sharp as well.
\end{example}

\subsection{POV-measure with commuting range}

In this section, we keep our standing assumption that $L$ is a
$\sigma$-orthocomplete effect algebra, and $S$ is an order
determining set of $\sigma$-additive states on $L$.

Let $\xi:(X,{\mathcal A})\to L$ and $\eta :(Y,{\mathcal B}) \to L$
be observables on $L$. Assume that $\eta$ is a smearing of $\xi$,
hence there is a (weak) Markov kernel $\nu(x,E):X\times {\mathcal
B}\to [0,1]$ such that for every $m\in S$, and $E\in {\mathcal B}$,
$$
m(\eta(E))=\int_X\nu_E(x)m(\xi(dx)).
$$

For $E$ fixed, $\nu_E:X\to [0,1]$ is a measurable function, and the
right hand side of the above equality is a mean value of the
observable $\nu_E(\xi)$ in the state $m$, so that for every
$\sigma$-additive state $m$,
$$
m(\eta(E))=m(\nu_E(\xi)).
$$

Define the observable $\Lambda_{\eta(E)}$ by
$\Lambda_{\eta(E)}\{1\}=\eta(E)$, $\Lambda_{\eta(E)}\{0\}=\eta(E)'$,
then we obtain, for all $m\in S$,
$$
m(\Lambda_{\eta(E)})=m(\nu_E(\xi)).
$$

\begin{theorem}\label{th:3} Let $\xi:(X,{\mathcal A})\to L$ and
$\eta :(Y,{\mathcal B})\to L$ be  observables on $L$,  such that
$\eta$ is a smearing of $\xi$ with a (weak) Markov kernel $\nu$ such
that for all Borel sets $E$, $\nu(x,E)\in \{ 0,1\}$ a.e. $m\circ
\xi$, $m\in S$. Then ${\mathcal R}(\eta)\subseteq {\mathcal
R}(\xi)$.

If $\eta$ and $\xi$ are sharp, and $\eta$ is real then also the
converse statement is true.
\end{theorem}
\begin{proof} Let $E\in {\mathcal B}$.
Under the hypotheses, we have for every $m\in S$,
\begin{eqnarray*}
m(\eta(E))&=&\int \nu(x,E)m(\xi(dx))\\
&=&\int_{\{x: \nu(x,E)=1\}}\nu(x,E)m(\xi(dx))\\
&=&m(\xi(\{ x: \nu(x,E)=1\})).
\end{eqnarray*}
Since $S$ is order determining, we obtain
$\eta(E)=\xi(\nu_E)^{-1}(\{ 1\})\in {\mathcal R}(\xi)$.

We note that in this case we have $\Lambda_{\eta(E)}=\nu_E(\xi)$.

Let both $\eta$ and $\xi$ be sharp, and assume that ${\mathcal
R}(\eta)\subseteq {\mathcal R}(\xi)$. Since the range of a sharp
observable is a Boolean $\sigma$-algebra \cite{DvPu1}, if $\eta$ is
real, we can apply \cite[Theorem 1.4]{Var}, to obtain that there is
a measurable function $f:X\to {\mathbb R}$ such that
$$
\eta(E)=\xi(f^{-1}(E)
$$
for all Borel sets $E\subseteq {\mathbb R}$, and the function $f$ is
unique up to a $\xi$-null set. Putting $\nu(x,E)=\chi_{f^{-1}(E)}$,
we obtain
$$
m(\eta(E))=m(\xi(f^{-1}(E))=\int \nu(x,E)m(\xi(dx))
$$
for all states $m\in S$. Hence $\eta$ is a smearing of $\xi$ with a
Markov kernel $\nu(x,E)=\chi_{f^{-1}(E)}(x)\in \{ 0,1\}$.
\end{proof}

\begin{theorem}\label{th:4} On the effect algebra ${\mathcal E}(H)$
of a separable $H$, an  observable (POV-measure) is a smearing of a
sharp  observable (PV-measure) if and only if the range ${\mathcal
R}(\eta)$ of $\eta$ consists of mutually commuting effects.

Moreover, a sharp real observable $\eta$ is a smearing of a sharp
observable $\xi$ if and only if ${\mathcal R}(\eta)\subseteq
{\mathcal R}(\xi)$, equivalently, if and only if $\eta$ is a
function of $\xi$.
\end{theorem}
\begin{proof}
Let $\eta:(Y,{\mathcal B})\to {\mathcal E}(H) $ be a  POV- measure
that is a smearing of a PV-measure $\xi:(X,{\mathcal A})\to
{\mathcal E}(H)$. Then for every set $E$, and every state $m$,
$m(\eta(E))=\int_X \nu(x,E)m(\xi(dx))=m(\nu_E(\xi))$, which implies
that $\eta(E)=\nu_E(\xi)$, where  $\nu_E(\xi)$ is a function of
$\xi$. It follows that all the spectral projections of the
self-adjoint operator $\eta(E)$ belong to ${\mathcal R}(\xi)$, and
this implies that ${\mathcal R}(\eta)$  consists of mutually
commuting effects.

Conversely, let the range of a POV measure $\eta$ consist of
commuting effects. By well known von Neumann theorem (see also
\cite{Var}), there exists a self-adjoint operator $V$ and Borel
measurable functions $f_E$ such that $\eta(E)=f_E(V)$. Then we have,
for every state $m$,
$$
m(\eta(E))=\int_X f_E(x) m(P^V(dx)),
$$
where $P^V$ is the spectral measure of $V$. Define
$\nu(x,E):=f_E(x)$. We will show that $\nu$ is a weak Markov kernel.

(i) Since $0\leq \eta(E) \leq 1$, $0\leq f_E(x)\leq 1$ on the
spectrum of $V$, hence $0\leq f_E\leq 1$ a.e. $m\circ P^V$ for all
$m$.

(ii) $f_Y(V)=\eta(Y)=1$ implies that $\int_X\nu(x,Y)m(P^V(dx))=1$,
and as $0\leq \nu(x,Y)\leq 1$, we get $\nu(x,Y)=1$ a.e. $m\circ P^V$
for all $m$. Similarly we show that $\nu(x,\emptyset)=0$ a.e.
$m\circ P^V$ for all $m$.

(iii) Let $E=\bigcup_{i=1}^{\infty}E_i$, where $E_i\cap
E_j=\emptyset$ whenever $i\neq j$. From
\begin{equation}\label{eq:eta}
\eta(E)= \sum_{i=1}^{\infty} \eta(E_i)
\end{equation}
 (the convergence in weak sense), we obtain that
\begin{equation}\label{eq:eta1}
\int \nu(x,E)P^V(dx)=\sum_{i=1}^{\infty} \int \nu(x,E_i)P^V(dx).
\end{equation}
Moreover, for every $n$,
$\eta(\bigcup_{i=1}^nE_i)=\sum_{i=1}^n\eta(E_i)$ entails
$f_{\bigcup_{i=1}^nE_i}(V)=\sum_{i=1}^nf_{E_i}(V)=(\sum_{i=1}^n
f_{E_i})(V)$, which entails that
$f_{\bigcup_{i=1}^n}(x)=\sum_{i=1}^n f_{E_i}(x)$ on the spectrum of
$V$. From this we derive that $0\leq
\nu(x,\bigcup_{i=1}^nE_i)=\sum_{i=1}^n\nu(x,E_i)\leq 1$ for all $n$.
Therefore $\sum_{i=1}^{\infty}\nu(x,E_i)$ exists, and (\ref{eq:eta})
yields that
$$
f_E(V)=\sum_{i=1}^{\infty}f_{E_i}(V),
$$
whence $f_E(x)=\sum_{i=1}^{\infty}f_{E_i}(x)$ on the spectrum of
$V$. We conclude that $\nu(x,E)=\sum_{i=1}^{\infty}E_i$ a.e. $m\circ
P^V$ for every state $m$. This concludes the proof that $\nu$ is a
weak Markov kernel, and $\eta$ is a smearing of $\xi:= P^V$.

Let both $\xi$ and $\eta$ be sharp, and let $\eta$ be real. If
${\mathcal R}(\eta)\subseteq {\mathcal R}(\xi)$,
 Theorem \ref{th:3} implies that $\eta$ is a smearing of $\xi$.

Conversely, assume that $\eta$ is a smearing of $\xi$ with a (weak)
Markov kernel $\nu$. For every state $m$, and every $E\in {\mathcal
B}({\mathbb R})$, $m(\eta(E))=\int
\nu(x,E)m(\xi(dx))=m(\nu_E(\xi))$. We may also write
$m(\Lambda_{\eta(E)})=m(\nu_E(\xi))$ for every $m$, where
$\Lambda_{\eta(E)}$ is the $0-1$ observable associated with
$\eta(E)$, which yields $\Lambda_{\eta(E)}=\nu_E(\xi)$. Then
$\eta(E)=\lambda_{\eta(E)}\{1\}=\xi(\nu_E^{-1}\{ 1\})\in {\mathcal
R}(\xi)$. We obtained that ${\mathcal R}(\eta)\subseteq {\mathcal
R}(\xi)$, equivalently, that $\eta=f(\xi)$ for a measurable function
$f$.
\end{proof}

\subsection{Some examples of minimal observables}
\begin{example}\label{ex:bud} Let $L$ be any $\sigma$-orthocomplete
effect algebra, and let $a_1,a_2,\ldots, a_n$ be elements of $L$
such that $\oplus_{i\leq n} a_i=1$. Choose real numbers
$r_1,r_2,\ldots,r_n$. Then we may construct a (real) observable
$\xi$ on $L$ by putting $\xi(E)=\oplus_{\{i:r_i\in E\}} a_i$, $E\in
{\mathcal B}({\mathbb R})$. We clearly have $\xi(\{ r_i\})=a_i$,
$i\leq n$, and $\{ r_1,r_2,\ldots,r_n\}$ is the spectrum of $\xi$.

Now let $L={\mathcal E}(H)$, where $H$ is a finite dimensional
Hilbert space. Let $A_1,A_2,\ldots,\\A_n$ be effects in ${\mathcal
E}(H)$ such that $\sum_{i\leq n}A_i=1$. That is,
$A_1,A_2,\ldots,A_n$ is a resolution of unity in ${\mathcal E}(H)$.
Let $\eta$ be a real observable on ${\mathcal E}(H)$ such that
$\eta(E)=\sum_{\{i: \alpha_i\in E\}}A_i$, $E\in {\mathcal
B}({\mathbb R})$, where $\alpha_i, i\leq n$ are real numbers.
Clearly, $\eta(\alpha_i)=A_i$, $i\leq n$, and $\{ \alpha_i:i\leq
n\}$ is the spectrum of $\eta$. Notice that every POV measure $\eta$
on a finite dimensional Hilbert space $H$ is of this type, and
$\eta(\{ \alpha_i\})$, $i\leq n$, are atoms of the range ${\mathcal
R}(\eta)$ of the observable $\eta$.

Since every $A_i, i\leq n$ is a self adjoint operator on $H$, it has
a spectral decomposition $A_i=\sum_{j=1}^{k_i}a_{i_j}P_{i_j}$, where
$P_{i_j}$'s are one dimensional projections with
$\sum_{j=1}^{k_i}P_{i_j}=1$, and $0\leq a_{i_j}\leq 1$ are
eigenvalues of $A_i$ (not necessarily all different). The elements
$B_{i_j}:=a_{i_j}P_{i_j}$ are effects in ${\mathcal E}(H)$. Owing to
$\sum_{i\leq n}A_i=1$, we have $\sum_{i\leq
n}\sum_{j=1}^{k_i}B_{i_j}=1$. By the first paragraph, we may choose
real numbers $(\beta_{i_j})_{i_j}$ and construct an observable $\xi$
such that $\xi(\beta_{i_j})=B_{i_j}$, and more generally,
$\xi(E)=\sum_{\{i_j:\beta_{i_j}\in E\}} B_{i_j}$, $E\in {\mathcal
B}({\mathbb R})$.

Define $f:{\mathbb R}\to {\mathbb R}$ by $f(\beta_{i_j})=\alpha_i$,
$i=1,\ldots, n$, $j=1,\ldots,k_i$ and $f(r)=0$ if $r\neq
\beta_{i_j}$ for all $i_j$. Since the range of $f$ is finite, it is
measurable. Moreover, $f^{-1}(\alpha_i)= \{ \beta_{i_j},
j=1,2,\ldots, k_i\}$. Therefore,
$\eta(\alpha_i)=A_i=\sum_{j=1}^{k_i}B_{i_j}
=\sum_{j=1}^{k_i}\xi(\beta_{i_j})=\xi(f^{-1}(\alpha_i)$. Hence
$\eta=f\circ \xi$, and hence $\eta$ is a smearing of $\xi$.

We have the following conclusion: if the range of a POV observable
$\eta$ contains atoms with rank greater than 1, then there is an
observable $\xi$ with $\xi \preceq \eta$. It follows that $\eta$ is
not minimal. The converse statement is also proved in \cite{Bud}.
Now we will rewrite it in our setting.

Assume that $\eta$ is a POV measure with the spectrum $\{
y_j\}_{j\leq n}$ such that $\eta(y_j)$ for every $j\leq n$ is an
effect of rank one, that is, a multiple of a one-dimensional
projection.  Assume that $\eta$ is a smearing of a POV $\xi$. This
entails, for every $j$,
\begin{eqnarray*}
\eta(y_j) &=& \sum_i \nu(x_i,y_j)\xi(x_i)\\
&=& \sum_{i\in \ell(j)}\nu(x_i,y_j) \xi(x_i),
\end{eqnarray*}
where $i\leq m$ for some $m\in {\mathbb N}$ and we put, for every
$j$, $\ell(j)=\{ i: \nu(x_i,y_j)\neq 0\}$. Observe that  for every
$i$, $\sum_j\nu(x_i,y_j)=1$, since $\nu(x_i,.)$ is a probability
measure.

Since $\eta(y_j)$ is rank one, we have $\xi(x_i)=\beta_i \eta(y_j)$
$\forall i\in \ell(j)$, with $0<\beta_i\leq 1$. This yields
\begin{equation}\label{eq:clean1}
 \eta(y_j)=\sum_{i\in \ell(j)}\nu(x_i,y_j)\beta_i\eta(y_j).
\end{equation}
Define $\alpha_i^j:=\nu(x_i,y_j)\beta_i$, then (\ref{eq:clean1})
implies $\sum_{i\in \ell(j)}\alpha_i^j=1$,
$\alpha_i^j\eta(y_j)=\nu(x_i,y_j)\xi(x_i)$, and from
$$
\sum_j\nu(x_i,y_j)\xi(x_i)=\xi(x_i)
$$
we get, putting $\bar{\nu}(y_j,x_i):=\alpha_i^j$, if $i\in \ell(j)$,
$\bar{\nu}(y_j,x_i):=0$ otherwise,
\begin{equation}\label{eq:clean2}
\xi(x_i)=\sum_j\bar{\nu}(y_j,x_i)\eta(y_j),
\end{equation}
which shows that $\xi$ is a smearing of $\eta$. This shows that
$\xi\preceq \eta$ implies  $\eta \sim \xi$, i.e. $\eta$ is minimal.
\end{example}

\begin{example}\label{ex:he} Let $L$ be a $\sigma$-orthocomplete effect algebra,
$(\Omega,{\mathcal A})$ a measurable space. Let ${\mathcal
O}(\Omega, {\mathcal A}, L)$ denote the set of all observables on
$L$ with the value space $(\Omega, {\mathcal A})$. Let ${\mathcal
O}\subseteq {\mathcal O}(\Omega, {\mathcal A}, L)$. We say that an
observable $\xi$ is {\it minimal in} ${\mathcal O}$  (or {\it
${\mathcal O}$-clean}), if for any $\eta \in {\mathcal O}$, the
condition $\eta \preceq \xi$ implies that $\eta \sim \xi$.

Recall that any element $a$ in an effect algebra $L$ defines an
observable $\xi^a$ with the outcome space $\Omega =\{ 0,1\}$ by
$$
\xi^a(\{ 1\})=a, \ \xi^a(\{ 0\})=a'.
$$
Observables of this type are called {\it 1-0-observables}.

In \cite{He}, minimality in the class  ${\mathcal O}(\{ 0,1\},H)$ of
the 1-0 observables on the effect algebra ${\mathcal E}(H)$ is
considered.  In accordance with \cite{He}, the 1-0 observable
corresponding to $A\in {\mathcal E}(H)$ will be denoted by $E^A$. We
recall that the set ${\mathcal O}(\{ 0,1\}, H)$ is convex, and
$$
\lambda E^A+(1-\lambda)E^B=E^{\lambda A+(1-\lambda)B}
$$
whenever $\lambda \in [0,1]$.

\begin{proposition}\label{pr:he} {\rm \cite{He}} Let $A,B\in
{\mathcal E}(H)$ and let $E^A,E^B$ be the corresponding 1-0
observables. Then $E^B\preceq E^A$ if and only if there are numbers
$s,t\in [0,1]$ such that $A=tB+sB'$.
\end{proposition}

As a consequence of Proposition \ref{pr:he} we obtain that $E^A\sim
E^B$ iff $A=B$ or $A=B'$.

\begin{proposition}\label{pr:he1} {\rm \cite[Prop. 3]{He}}
Let $A\in {\mathcal E}(H)$. The observable $E^A$ is minimal in
${\mathcal O}(\{ 0,1\},H)$ if and only if $\mid \mid A\mid \mid=\mid
\mid A'\mid \mid =1$.
\end{proposition}

We note that the extreme elements in the convex set ${\mathcal
E}(H)$ are projection operators \cite[Lemma 2.3]{Dav}. According to
proposition \ref{pr:he1}, for every projection $P$, the
corresponding 1-0 observable $E^P$ is minimal in ${\mathcal O}(\{
0,1\},H)$, but there are also other minimal observables. However, in
accordance with Example \ref{ex:bud}, if $\dim H\geq 3$, then the
observable $E^A$ for any effect $A$ is not minimal in the set of all
real observables on ${\mathcal E}(H)$.
\end{example}

\section{Smearings of observables on effect algebras with the
(E)-property}

Notice that if ${\mathcal S}$ is an order determining system of
states on an effect algebra $L$, then by replacing ${\mathcal S}$ by
its ($\sigma$)-convex hull $Conv({\mathcal S})$, we may always
assume that ${\mathcal S}$ is a $(\sigma$)-convex set.

Let $L$ be a $\sigma$-orthocomplete effect algebra with an order
determining system ${\mathcal S}$ of $\sigma$-additive states,
$(X,{\mathcal F})$ be a measurable space. Every observable $\xi
:{\mathcal F}\to L$ can be characterized by a mapping $\Phi_{\xi}
:{\mathcal S} \to M_1^+(X,{\mathcal F})$ defined by
\begin{equation}\label{eq:phi}
\Phi_{\xi}(m)(F)=m\circ \xi(F), \ m\in {\mathcal S}, F\in {\mathcal
F}.
\end{equation}
Here $F\to m\circ \xi(F)=\Phi_{\xi}(m)(F),\ F\in {\mathcal F}$ is
the probability distribution of the observable $\xi$ in the state
$m$.

We will try to find conditions under which to  given observable and
Markov kernel there exists a fuzzy version. For the sake of
simplicity, we will concentrate to real observables.

The following definitions were introduced in \cite{DDD}. Let $S\neq
\emptyset$ be a convex set. A mapping $f:S\times {\mathcal
B}({\mathbb R}) \to [0,1]$ such that
\begin{enumerate}
\item[(i)] given $m\in S$, $f(m,.)$ is a probability measure on
${\mathcal B}({\mathbb R})$,
\item[(ii)] for any $E\in {\mathcal B}({\mathbb R})$, $f(\lambda
m_1+(1-\lambda)m_2,E)=\lambda f(m_1,E)+(1-\lambda)f(m_2,E)$ whenever
$\lambda \in [0,1]$ and $m_1,m_2\in S$
\end{enumerate}
is said to be a {\it $\sigma$-effect function} on $S$.

\begin{definition}\label{de:1} Let $L$ be an effect algebra. We say
that a convex system $S$ of states on $L$ has the (E)-{\it property}
({\rm E} as for existence) if, given a $\sigma$-effect function $f$
on $S$, for any $E\in {\mathcal B}({\mathbb R})$ there exists an
element $\xi(E)\in L$ such that
\begin{enumerate}
\item[{\rm(E)}] $f(m,E)=m(\xi(E)), m\in S$.
\end{enumerate}
\end{definition}

It was shown in \cite{DDD} that the set of all $\sigma$-additive
states on the effect algebra ${\mathcal E}(H)$ has the (E)-property.

\begin{theorem}\label{th:1}  Let $L$ be a $\sigma$-orthocomplete
effect algebra and let $S$ be a convex order determining system of
$\sigma$-additive states on $L$ which has the (E)-property. Then for
every effect function $f$ on $S$, the mapping $E\mapsto \xi(E)$ from
${\mathcal B}({\mathbb R})\to L$ is an observable on $L$. Moreover,
$m\mapsto f(m,.)=\Phi_{\xi}(m)$.
\end{theorem}
\begin{proof}
Since $S$ is order determining, the element $\xi(E)$ is uniquely
defined by property (E). Moreover, $\xi({\mathbb R})=1$. Let
$(E_i)_{i=1}^{\infty} \subseteq {\mathcal B}({\mathbb R})$ be such
that $E_i\cap E_j=\emptyset$ whenever $i\neq j$. Then for any $i\neq
j$,
\begin{eqnarray*} 1&\geq& s(\xi(E_i\cup E_j))=f(m,E_i\cup E_j)\\
&=&f(m,E_i)+f(m,E_j)=m(\xi(E_i))+m(\xi(E_j)),
\end{eqnarray*}
which implies that $\xi(E_i)\perp \xi(E_j)$ and $\xi(E_i\cup
E_j)=\xi(E_i)\oplus \xi(E_j)$. By induction we prove that
$\xi(\bigcup_{i=1}^nE_i)=\bigoplus_{i=1}^n\xi(E_i)$. Put
$E:=\bigcup_{i=1}^{\infty}E_i$. Then for every $m\in S$,
\begin{eqnarray*}
m(\xi(E))&=&f(m,\bigcup_{i=1}^{\infty}E_i)=\sum_{i=1}^{\infty}f(m,E_i)\\
&=& \lim_{n\to\infty}\sum_{i=1}^nf(m,E_i)=\lim_{n\to
\infty}\sum_{i=1}^nm(\xi(E_i))\\
&=&\lim_{n\to\infty}m(\bigoplus_{i=1}^n\xi(E_i))=
m(\bigoplus_{i=1}^{\infty}\xi(E_i)),
\end{eqnarray*}
where the last equality holds owing to $\sigma$-additivity of $m$.
Since $S$ is order determining, we obtain
$\xi(E)=\bigoplus_{i=1}^{\infty}\xi(E_i)$.
\end{proof}

\begin{theorem}\label{th:2} Let $S$ be a convex order determining
system of $\sigma$-additive states on a $\sigma$-orthocomplete
effect algebra $L$ such that $S$ has (E) property. Then, given a
real observable $\xi$ on $L$, and a Markov kernel $\nu:{\mathbb
R}\times {\mathcal B}({\mathbb R})\to [0,1]$ there is a fuzzy
version $\eta$ of $\xi$.
\end{theorem}
\begin{proof} Let $\xi$ and $\nu$ be given. For every $m\in S$ and
$E\in {\mathcal B}({\mathbb R})$, define
\begin{equation}\label{eq:10} f(m,E):=\int_{\mathbb
R}\nu_E(x)m(\xi(dx)).
\end{equation}
We will prove that $f:S\times {\mathcal B}({\mathbb R})\to [0,1]$ is
an effect function. (i):Let $m\in S$ be fixed, and let
$(E_i)_{i=1}^{\infty}$ be a sequence of disjoint sets from
${\mathcal B}({\mathbb R})$. Put $E=\bigcup_{i=1}^{\infty}E_i$. Then
$\nu_E(x):=\nu(x,E)=\sum_{i=1}^{\infty}\nu_{E_i}(x)$, and by
additivity of the integral, for every $n\in {\mathbb N}$,
$$
\int_{\mathbb R}\nu(x,\bigcup_{i=1}^nE_i)m(\xi(dx))
=\sum_{i=1}^n\int_{\mathbb R}\nu(x,E_i)m(\xi(dx)).
$$
Since $\nu_{\bigcup_{i=1}^n E_i}\nearrow \nu_E$ pointwise, we have
$$
\lim_{n\to \infty}\int_{\mathbb
R}\nu(x,\bigcup_{i=1}^nE_i)m(\xi(dx))=\int_{\mathbb
R}\nu_E(x)m(\xi(dx)),
$$
hence $f(m,E)=\sum_{i=1}^{\infty}f(m,{E_i})$.

(ii): Let $m=\alpha m_1+(1-\alpha)m_2$, then for every $E\in
{\mathcal B}({\mathbb R})$, $m(\xi(E))=\alpha
m_1(\xi(E))+(1-\alpha)m_2(\xi(E))$. Therefore
\begin{eqnarray*}
f(m,E)&=& \int_{\mathbb R}\nu_E(x)m(\xi(dx))\\
&=&\int_{\mathbb R}\nu_E(x)(\alpha m_1(\xi(dx))+(1-\alpha)m_2(\xi(dx)))\\
&=&\alpha \int_{\mathbb
R}\nu_E(x)m_1(\xi(dx))+(1-\alpha)\int_{\mathbb R}\nu_E(x)m_2(\xi(dx))\\
&=& \alpha f(m_1,E)+(1-\alpha)f(m_2,E).
\end{eqnarray*}

This proves that $f$ is an effect function. Then the (E) property
entails that there is an observable $\eta$ such that for every $m\in
S$, $E\in {\mathbb R}$, $m(\eta(E))=\int_{\mathbb
R}\nu(x,E)m(\xi(dx))$, that is, $\eta$ is a fuzzy version of $\xi$.
\end{proof}

\section{Stochastic operators and Markov kernels}

At the beginning, we introduce some notations. Let
$M(\Omega,{\mathcal A})$ denote the vector space of all complex
measures on $(\Omega, {\mathcal A})$. Then $M(\Omega, {\mathcal A})$
is a Banach space with the total variation norm $\mid \mid \mu \mid
\mid= \mid \mu \mid(\Omega)$.

Let $\mu$ be a $\sigma$-finite measure on $(\Omega, {\mathcal A})$,
we denote by $L(\mu)$ the subspace in $M(\Omega, {\mathcal A})$
generated by all $P\in M_1^+(\Omega, {\mathcal A})$ such that $P\ll
\mu$. The space $L(\mu)$ can be identified with $L_1(\Omega,
{\mathcal A}, \mu)$, by extension of the map $P\mapsto
\frac{dP}{d\mu}$. If ${\mathcal P}\subset M_1^+(\Omega, {\mathcal
A})$, then we denote by $L({\mathcal P})$ the subspace generated by
$\bigcup_{P\in {\mathcal P}}L(P)$.

A {\it stochastic operator} is an affine map
$$
T: \mathcal M \to M_1^+(\Omega_1, {\mathcal A}_1).
$$
where $\mathcal M$  is a convex subset in $M_1^+(\Omega, {\mathcal A})$.
Any stochastic operator can be extended to a positive, norm - preserving
map from the Banach subspace in  $M(\Omega, {\mathcal A})$
generated by $\mathcal M$, to  $M(\Omega_1, {\mathcal A}_1)$.

\begin{example}\label{ex:stoop}
{\bf 1.} Let ${\mathcal A}_1\subset {\mathcal A}$ be a
sub-$\sigma$-algebra. Then the restriction map
$$
T_{{\mathcal A}_1}: M_1^+(\Omega, {\mathcal A})\to M_1^+(\Omega,
{\mathcal A}_1), \, P\mapsto P/{\mathcal A}_1
$$
is a stochastic operator.

{\bf 2.} Let $F:(\Omega, {\mathcal A})\to (\Omega_1, {\mathcal
A}_1)$ be a measurable map. Then $F$ defines the stochastic operator
$$
T^F:M_1^+(\Omega, {\mathcal A})\to M_1^+(\Omega_1, {\mathcal A}_1),
\, P\mapsto P^F,
$$
where $P^F$ is the distribution of $F$ under $P$, that is,
$$
P^F(B)=P(F^{-1}(B)), \, B\in {\mathcal A}_1.
$$

{\bf 3.} Let $\nu :\Omega \times {\mathcal A}_1\to {\mathbb R}$ be a
weak Markov kernel with respect to ${\mathcal P}$. Then
$T_{\nu}:P\mapsto \nu(P)$ defines a stochastic operator
$T_{\nu}:L({\mathcal P})\to M(\omega_1, {\mathcal A}_1)$.
\end{example}

A stochastic operator is called a {\it statistical map} if there is
a Markov kernel such that $T=T_{\nu}$.
Note that operators $T_{{\mathcal A}_1}$ and $T_F$ in Example
\ref{ex:stoop} are given by Markov kernels. Indeed, if we put

\begin{equation}\label{eq:stoop}
\nu_{{\mathcal A}_1}(\omega,B)=\chi_B(\omega), \, \, \nu_F(\omega,
B)=\chi_{F^{-1}(B)}(\omega),
\end{equation}
then $T_{{\mathcal A}_1}=T_{\nu_{{\mathcal A}_1}}$, and
$T_F=T_{\nu_F}$.

\begin{proposition}\label{pr:Bug} {\rm \cite{BHS, Bu}} A stochastic
operator $T:M_1^+(\Omega, {\mathcal A}) \to M_1^+(\Omega_1,
{\mathcal A}_1)$ is a statistical map if and only if for every $B\in
{\mathcal A}_1$ there is an ${\mathcal A}$-measurable function $f_B:
\Omega \to [0,1]$ such that
\begin{equation}\label{eq:Bug}
\int_B(TP)(d\omega_1)= \int_{\Omega} f_B(\omega)P(d\omega)
\end{equation}
for every $P\in M_1^+(\Omega, {\mathcal A})$.
\end{proposition}
\begin{proof} If there is a Markov kernel $\nu$ such that
$T=T_{\nu}$, we put $f_B(\omega)=\nu(\omega, B)$.

Conversely, $\nu(\omega, B):=f_B(\omega)$ is a
 Markov kernel. Indeed, for every $\omega^*\in \Omega$ let $\delta_{\omega^*}$
denote the corresponding Dirac measure. Equation (\ref{eq:Bug})
implies that
$T\delta_{\omega^*}(B)=\int_{\Omega}f_B(\omega)\delta_{\omega^*}(d\omega)
=f_B(\omega^*)$, which immediately implies the desired result.
\end{proof}

Let $\xi :(X,{\mathcal F})\to L$ and $\eta :(Y,{\mathcal G})\to L$
be observables such that $\xi \preceq \eta$, and let $\nu :X\times
{\mathcal G}\to [0,1]$ be the corresponding confidence measure. Then
the equation
\begin{equation}\label{eq:preceq}
 m(\eta(G))=\int_X \nu(x,G)m(\xi(dx)),\ G\in {\mathcal
G},\ m\in {\mathcal S}
\end{equation}
can be rewritten in the form
\begin{equation}\label{eq:stoch}
\Phi_{\eta}=T_{\nu}\circ \Phi_{\xi},
\end{equation}
where $T_{\nu}$ is the statistical map corresponding to $\nu$, and
$\Phi_{\xi}$ is defined by (\ref{eq:phi}).

In general, there is a little hope for a stochastic operator
$T:M(\Omega, {\mathcal A})\to M(\Omega_1, {\mathcal A}_1)$ to be
given by a Markov kernel (see e.g. \cite{BHS}). However, for
stochastic operators defined on $L(\mu)$ it is often the case.

Let us recall that the measurable space $(X,{\mathcal B})$ is a {\it
standard Borel space} if $X$ is a complete separable metrizable
(Polish) space and ${\mathcal B}$ is the Borel $\sigma$-algebra over
$X$.

\begin{theorem}\label{th:stand} {\rm \cite{Str}}. Let $\mu$ be a $\sigma$-finite
measure on $(\Omega, {\mathcal A})$ and $(X,{\mathcal B})$ be a
standard Borel space. Let $T: L(\mu)\to M(X,{\mathcal B})$ be a
stochastic operator. Then there is a Markov kernel $\nu:\Omega \times
{\mathcal B}\to [0,1]$ such that $T=T_\nu/L(\mu)$.
\end{theorem}

Let $\xi$ be an $(\Omega, {\mathcal A})$-observable on $L$, and let
$m_0$ be a faithful $\sigma$-additive state on $L$. Then for every
$m\in {\mathcal S}$, we have $m\circ{\xi}\ll m_0\circ{\xi}$. By
Theorem \ref{th:stand}, to every stochastic operator
$$
T: L(m_0\circ{\xi}) \to M(X,{\mathcal B}),
$$
where $(X,{\mathcal B})$ is a standard Borel space, there is a
Markov kernel $\nu:\Omega \times B\to [0,1]$ such that
$T=T_{\nu}/L(m_0\circ{\xi})$.

\subsection{Coarse graining}

We keep our assumption that $L$ is a $\sigma$-complete effect
algebra and ${\mathcal S}$ is an order determining system of
$\sigma$-additive states on $L$.

\begin{definition}\label{de:cg} Let $\xi$ and $\eta$ be observables
on $L$ with value spaces $(X,{\mathcal F})$ and $(Y, {\mathcal G})$,
respectively. We say that $\eta$ is a {\it coarse graining} of $\xi$
if there is a stochastic operator $T:M_1^+(X,{\mathcal F})\to
M_1^+(Y, {\mathcal G})$ such that $\Phi_{\eta}=T\circ \Phi_{\xi}$,
where $\Phi_{\xi}$ is defined by (\ref{eq:phi}).
\end{definition}

Our previous discussion shows that if $\xi \preceq \eta$, then
$\eta$ is a coarse-graining of $\xi$. If $\eta$ is a coarse graining
of $\xi$, and  conditions of Theorem \ref{th:stand} are satisfied,
then $\xi \preceq \eta$ holds.

In particular, if $L={\mathcal E}(H)$ ($H$ separable), then for any
observable $\xi :(\Omega, {\mathcal A})\to L$ and any faithful state
$m_0$, every stochastic operator $T: L(m_{0\xi})\to M(X,{\mathcal
B})$ (where $(X,{\mathcal B})$ is a standard Borel space) defines a
Markov kernel $\nu$, and hence a fuzzy version $\eta$ of $\xi$ with
the confidence measure $\nu$. If $(X,{\mathcal B})= ({\mathbb R},
{\mathcal B}({\mathbb R}))$, then $\eta$ is a real observable.

\section{Application of the classical theory of experiments to
quantum observables}

Representations of observables by probability distributions enable
us to apply some results from the classical theory of experiments to
quantum observables.

A  (statistical) {\it experiment} (or {\it model}) is a triple
$X=(\Omega, {\mathcal A}, {\mathcal P})$, where $(\Omega, {\mathcal
A})$ is a measurable space and ${\mathcal P}$ is a nonempty family
of measures in $M_1^+(\Omega, {\mathcal A})$. $(\Omega, {\mathcal
A})$ is called the {\it sample space} of the experiment $X$.

\subsection{f-divergence}

Let $P,Q\in M_1^+(\Omega, {\mathcal A})$. Recall that a {\it
Lebesgue decomposition} of $P$ with respect to $Q$ is any pair
$(f,N)$,such that $f:\Omega \to {\mathbb R}$ is measurable, $f\geq
0$, $N\in {\mathcal A}$, $Q(N)=0$, and
$$
P(A)=\int_A fdQ + P(A\cap N), \, A\in {\mathcal A}.
$$
The function $f$ is called the {\it likelihood ratio} of $P$ with
respect to $Q$ and denoted by $f=dP/dQ$. For example, if $P\ll Q$
and $(dP/dQ, N)$ is a Lebesgue decomposition, then $P(N)=0$ and
$dP/dQ$ is the Radon-Nikodym derivative. If both $P$ and $Q$ are
dominated by a $\sigma$-finite measure $\mu$ and $p=dP/d\mu$,
$q=dQ/d\mu$, $N=\{ q=0\}$, then $(p/q, N)$ is a Lebesgue
decomposition of $P$ w.r. $Q$.

Let $P,Q\in M_1^+(\Omega,{\mathcal A})$ and let $(dP/dQ,N)$ be a
Lebesgue decomposition. Let $f:[0,\infty) \to {\mathbb R}$ be a
convex function. We define the {\it $f$-divergence} of $P$ with
respect to $Q$ by \cite{Hey}
\begin{equation}\label{eq:fdiv}
D_f(P,Q)=\int_{\Omega}f(\frac{dP}{dQ})dQ + P(N)f_{\infty},
\end{equation}
where $f_{\infty}=\lim_{x\to \infty}\frac{f(x)}{x}$.
It can be seen that $D_f$ does not depend on the choice of the
Lebesgue decomposition.

The $f$-divergence of $P$ with respect to $Q$ is a generalization of
the total variation of $P-Q$. Choosing $f(u):=\mid u-1\mid$ for all
$u\in {\mathbb R}^+$ we obtain $D_f(P,Q)=\newline \mid \mid P-Q\mid
\mid$. More generally, if $f$ is a strictly convex function,
satisfying $f(1)=0$, then $D_f(P,Q)\ge 0$ for all $P$, $Q$  and
$D_f(P,Q)=0$ if and only if $P=Q$. In this sense, $D_f$ can be seen
as a quasi-distance in $M_1^+(\Omega,\mathcal A)$.

For  example, if $f(x)=-\log(x)$, then $D_f(P,Q)$ is the well-known
$I$-divergence (Kullback - Leibler divergence, relative entropy)
$$
I(P,Q)=\int_{\Omega}(\log q-\log p)qd\mu,
$$
here $\mu$ is a dominating measure and $p=dP/d\mu$, $q=dQ/d\mu$.

Another example is the Hellinger distance
$$
H(P,Q)=\frac{1}{2}\int_{\Omega} (p^{1/2}-q^{1/2})^2 d\mu
$$
obtained by the choice $f(x)=f_H(x)=(1-x^{1/2})$. For more
examples and facts about $f$-divergences, see \cite{LV}.

Let $\xi \preceq \eta$. The relation (\ref{eq:9}) implies that if
for two states $m_1,m_2$ we have $m_1\circ \xi=m_2\circ \xi$, then
also $m_1\circ \eta=m_2\circ \eta$ holds. That is, the discerning
power of $\xi$ with respect to states is greater than that of
$\eta$. A strengthening of this result is given by the following
{\it monotonicity theorem}. In \cite{Str, Hey}, it is proved for
Markov kernels, but the proof works also for weak Markov kernels.

\begin{theorem}\label{th:mon}
 Let $P,Q\in M_1^+(\Omega, {\mathcal A})$ and let $\nu :\Omega
 \times {\mathcal A}_1\to [0,1]$ be a weak Markov kernel with respect to
 $\{ P,Q\}$. Let $f:[0,\infty)\to {\mathbb R}$ be a convex function.
 Then
 $$
 D_f(P,Q)\geq D_f(\nu(P),\nu(Q)).
 $$
 \end{theorem}

\subsection{Sufficient Markov kernels}

Let $(\Omega,{\mathcal A},{\mathcal P})$ be an experiment,
$(\Omega_1, {\mathcal A}_1)$ a measurable space and $\nu:\Omega
\times {\mathcal A}_1\to [0,1]$ a Markov kernel. According to
(\ref{eq:mk}), $\nu$ assigns to every $P\in {\mathcal P}$ a measure
$\nu(P) \in M_1^+(\Omega_1, {\mathcal A}_1)$ by
$$
\nu(P)(A_1)=\int_{\Omega}\nu(x,A_1)P(dx).
$$
Note that for $P,Q\in M_1^+(\Omega,\mathcal A)$, $Q\ll P$ implies $\nu(Q)\ll
\nu(P)$. Indeed, if $B\in {\mathcal A}_1$
is such that $\nu(P)(B)=\int \nu(\omega,B)dP(\omega)=0$, then,
since $\nu(\omega,B)\geq 0$, we must have $\nu(\omega,B)=0, \, P$ a.e.
But then also $\int \nu(\omega,B)Q(d\omega)=\nu(Q)(B)=0$.

For a  measurable function $f:(\Omega,\mathcal A)\to [0,1]$ and
$P\in M_1^+(\Omega,\mathcal A)$, we define the measure $f\cdot P$ as
$$
f\cdot P(A): = \int_A fdP
$$
Then clearly $f\cdot P\ll P$, hence $\nu(f\cdot P)\ll \nu(P)$. Let
us define
$$
E^\nu_P(f):=d\nu(f\cdot P)/d\nu(P).
$$

\begin{definition}\label{de:suff} {\rm \cite[Definition 22.1]{Hey}}
Let $(\Omega, {\mathcal A}, {\mathcal P})$ be an experiment,
$(\Omega_1, {\mathcal A}_1)$ a measurable space and $\nu:\Omega
\times {\mathcal A}_1\to [0,1]$ a Markov kernel.
\begin{enumerate}
\item[(a)] $\nu$ is called {\it Blackwell sufficient} (for ${\mathcal
P}$) if there exists a kernel $\nu' :\Omega_1\times {\mathcal A}\to
[0,1]$ such that $\nu'(\nu(P))=P$ holds for all $P\in {\mathcal P}$.
\item[(b)] $\nu$ is said to be {\it sufficient} (for ${\mathcal P}$)
if to every $A\in {\mathcal A}$ there exists a measurable function
$g_A:(\Omega_1,{\mathcal A}_1)\to \mathbb R$, such that
\begin{equation}\label{eq:suff}
E_P^{\nu}(\chi_A)=g_A, \, \nu(P)\, a.e.\ \mbox{for all}\ P\in
{\mathcal P}.
\end{equation}
\end{enumerate}
\end{definition}

Clearly, for the observables $\xi, \eta$ on $L$, such that $\xi\preceq  \eta$,
we have $\xi \sim
\eta$ iff the corresponding Markov kernel $\nu$ is Blackwell
sufficient for the measures $m\circ \xi$, $m\in {\mathcal S}$.

\begin{remark} In classical statistics, sufficiency of a sub-$\sigma$-algebra
$\mathcal A_1$ (or a statistic) for an experiment means the
existence of common versions of the conditional probabilities
$P(A/\mathcal A_1)$ for all $P\in\mathcal P$. The above definition
is a generalization of this well-known notion: if $\mathcal
A_1\subset \mathcal A$ is a sub-$\sigma$-algebra and
$\nu=\nu_{\mathcal A_1}$, then $E_P^\nu(\chi_A)=P(A/\mathcal A_1)$.
\end{remark}

\begin{remark}\label{re:suff} Let us fix $P\in M_1^+(\Omega,\mathcal A)$.
Let us define a map $\nu_P':\Omega_1 \times {\mathcal A}\to \mathbb R$ by
$$
\nu_P'(\omega_1,A):=E_P^\nu(\chi_A)(\omega_1).
$$
Then we have
\begin{equation}\label{eq:suff}
\int_A\nu(\omega,B)P(d\omega)=\int_B
\nu_P'(\omega_1,A)\nu(P)(d\omega_1), \, A\in {\mathcal A}, B\in
{\mathcal A}_1.
\end{equation}
We prove that $\nu'_P$ is a weak  Markov kernel with respect to
$\nu(P)$.

By definition, we know that $\omega_1 \mapsto \nu_P'(\omega_1,A)$ is
measurable for all $A\in {\mathcal A}$. This shows (i).  Moreover,
$\nu_P'(\omega_1,A)\geq 0$, $\nu(P)$ a.e.. Moreover, let $B=\{
\omega_1: \nu_P'(\omega_1,A)>1\}$ and suppose that $\nu(P)(B)>0$.
Then by (\ref{eq:suff}),
$$
\nu(P)(B)<
\int_B\nu_P'(\omega_1,A)\nu(P)(d\omega_1)=\int_A\nu(\omega,B)P(d\omega)\leq
\nu(P)(B).
$$
It follows that $\nu(P)(B)=0$, whence $\nu_P'(\omega_1,A)\leq 1$,
$\nu(P)$ a.e., and (ii) is shown. We have
\begin{eqnarray*}
1&=& P(\Omega)=\int_{\Omega}\nu(\omega, \Omega_1)P(d\omega)\\
&=& \int_{\Omega_1}\nu_P'(\omega_1,\Omega)\nu(P)(d\omega_1)\\
&\Rightarrow &\nu_P'(\omega_1,\Omega)=1\, {\mbox a.e.}\ \nu(P)
\end{eqnarray*}
Similarly we show that $\nu_P'(\omega_1,\emptyset)=0\ {\mbox a.e.}\,
\nu(P)$ which proves (iii).
 Finally, let $\{ A_n\}$
be a sequence in ${\mathcal A}$, such that $A_n\cap A_m=\emptyset$
if $n\neq m$. Then for $B\in {\mathcal A}_1$,
\begin{eqnarray*}
\int_B\nu_P'(\omega_1, \bigcup_nA_n)\nu(P)(d\omega_1)&=&
\int_{\bigcup_nA_n}\nu(\omega,B)P(d\omega)=\sum_n\int_{A_n}\nu(\omega,B)P(d\omega)\\
&=&\int_B \sum_n \nu_P'(\omega_1, A_n)\nu(P)(d\omega_1)
\end{eqnarray*}
which proves (iv), so that $\nu'_P$ is indeed a weak Markov kernel
with respect to $\nu(P)$. By (\ref{eq:suff})
$$
\nu'_P(\nu(P)(A))=
\int_{\Omega_1}\nu_P'(\omega_1,A)\nu(P)(d\omega_1)=
\int_A\nu(\omega,\Omega_1)P(d\omega)=P(A).
$$

We see that sufficiency of the kernel $\nu$, in contrast with Blackwell
sufficiency, implies the existence of a {\it weak} Markov kernel $\nu'$, such
that $\nu'(\nu(P))=P$ holds for $P\in\mathcal P$.
\end{remark}

\subsection{Pairwise sufficiency}

We say that a subalgebra (Markov kernel) is {\it pairwise
sufficient} for ${\mathcal P}$, if it is sufficient for any pair $\{
P_1,P_2\}, P_1,P_2\in {\mathcal P}$. Clearly, a sufficient
subalgebra (Markov kernel) is pairwise sufficient.
We have the following characterization of pairwise sufficient Markov kernels.

\begin{theorem}\label{th:klc} {\rm \cite{LV, Hey}}{\rm (S. Kullback, R.A. Leibler, T.
Csisz\'ar)}. Let $(\Omega, {\mathcal A})$, $(\Omega_1,{\mathcal
A}_1)$ be measurable spaces, $\nu :\Omega\times {\mathcal A}_1\to
[0,1]$ a Markov kernel and $ P,Q\in M_1^+(\Omega,
{\mathcal A})$. Then the following are equivalent.
\begin{enumerate}
\item[{\rm(i)}] $\nu$ is sufficient for $\{ P,Q\}$.
\item[{\rm(ii)}] For any convex function $f$ on ${\mathbb R}^+$ one
has
\begin{equation}\label{eq:klc1} D_f(\nu(P),\nu(Q))=
D_f(P,Q).
\end{equation}
\item[{\rm(iii)}] There is a strictly convex function $f$ on ${\mathbb R}^+$
such that
\begin{equation}\label{eq:klc2}
D_f(\nu(P),\nu(Q))=D_f(P,Q)< \infty.
\end{equation}
\end{enumerate}
\end{theorem}
Note that we may take the Hellinger distance $H(P,Q)$ in (iii).

\subsection{Dominated families}

Let ${\mathcal P}\subset M_1^+(\Omega, {\mathcal A})$. We say that
${\mathcal P}$ is a {\it dominated family}, if there is a
$\sigma$-finite measure $\mu$ such that ${\mathcal P} \ll \mu$. If
this is the case, then we can find a finite measure $\mu_0$,
dominating ${\mathcal P}$. It is clear that if ${\mathcal P}\ll\mu$,
then we have $C({\mathcal P}) \ll \mu$, where
$$
C({\mathcal P})= \{ \sum_n\lambda_nP_n: \lambda_n \geq 0, \sum_n
\lambda_n =1, P_n\in {\mathcal P} \}.
$$
If we also have $\mu(A)=0$ whenever $P(A)=0$ for all $P\in {\mathcal
P}$, then we write ${\mathcal P} \sim \mu$.

\begin{lemma}\cite{HS}\label{le:HS} Let ${\mathcal P}$ be a dominated family.
Then there is a convex combination $P_0=\sum_n\lambda_n P_n$ of
elements of $P_n\in {\mathcal P}$, $n\in {\mathbb N}$, such that
${\mathcal P}\sim P_0$.
\end{lemma}

 The following theorem is well known.

\begin{theorem}\label{th:sufalg}{\rm \cite{ HS, Str}} Let ${\mathcal P} \subset
M_1^+(\Omega, {\mathcal A})$ be a dominated family and let $P_0\in C(\mathcal P)$ be such that $\mathcal P\sim P_0$.
Let ${\mathcal A}_1\subset {\mathcal A}$ be a
sub-$\sigma$-algebra. Then the following are equivalent.
\begin{enumerate}
\item[{\rm (i)}] $\mathcal A_1$ is sufficient for $\mathcal P$.
\item[{\rm (ii)}] $\mathcal A_1$ is  pairwise sufficient for ${\mathcal P}$.
\item[{\rm (iii)}] $\mathcal A_1$ is sufficient for the pair $\{P,P_0\}$ for every $P\in\mathcal P$.
\end{enumerate}
\end{theorem}

A similar statement holds also for Markov kernels. Since the proof
is not easy to find in the literature, we give it here. We will
first show that we can describe sufficient Markov kernels in terms
of sufficient subalgebras.

Let $\nu :\Omega \times {\mathcal A}_1\to [0,1]$ be a weak Markov
kernel with respect to ${\mathcal P}\subset M_1^+(\Omega,{\mathcal
A})$. For $P\in L({\mathcal P})\cap M_1^+(\Omega, {\mathcal A})$ we
define a probability measure $P\times \nu\in M_1^+(\Omega \times
\Omega_1, {\mathcal A}\otimes {\mathcal A}_1)$, by
\begin{equation}\label{eq:times}
P\times \nu(A\times B)=\int_A\nu(\omega, B)P(d\omega), \ A\in
{\mathcal A}, \, B\in {\mathcal A}_1.
\end{equation}
Note that we have
$P\times \nu (\Omega\times B)=\nu(B)$ and
$P\times \nu(A\times \Omega_1)=P(A)$ for $A\in \mathcal A$, $B\in \mathcal A_1$.

\begin{lemma}\label{le:sufalg} {\rm \cite{Hey}} Let
${\mathcal P}\subset M_1^+(\Omega, {\mathcal A})$ and let $\nu
:\Omega \times {\mathcal A}_1\to [0,1]$ be a Markov kernel. Then
$\nu$ is sufficient for ${\mathcal P}$ if and only if the
sub-$\sigma$-algebra ${\mathcal A}_0=\{ \emptyset, \Omega \}\otimes
{\mathcal A}_1\subset {\mathcal A}\otimes {\mathcal A}_1$ is
sufficient for  $\{ P\times \nu : P\in {\mathcal P}\}$.
\end{lemma}
\begin{proof} Let $\nu$ be sufficient and let $\nu'$ be the
corresponding weak Markov kernel, see Remark \ref{re:suff}. For $A\in {\mathcal A}, B\in
{\mathcal A}_1$, we define a function $f_{A\times B}:\Omega \times
\Omega_1 \to [0,1]$ by
$$
f_{A\times B}(\omega, \omega_1)=\nu'(\omega_1,A)\chi_B(\omega_1).
$$
It is clear that $f_{A\times B}$ is ${\mathcal A}_0$-measurable,
moreover, for $B_1\in {\mathcal A}_1$ and $P\in {\mathcal P}$,
\begin{eqnarray*}
\int_{\Omega \times B_1} f_{A\times B}d(P\times \nu)&=&\int_{B\cap
B_1}\nu'(\omega_1,A)\nu(P)(d\omega_1)=\int_A\nu(\omega, B\cap
B_1)P(d\omega)\\
&=&P\times \nu(A\times B\cap \Omega \times B_1).
\end{eqnarray*}

It follows that $f_{A\times B}$ is the common version of the
conditional probability  $f_{A\times B}=P\times \nu(A\times B/{\mathcal
A}_0)$, $P\times \nu$  a.e., for all $P\in {\mathcal P}$.

Conversely, suppose that ${\mathcal A}_0$ is sufficient for
$\{P\times \nu : P\in {\mathcal P}\}$ and let $f_{A\times B}=P\times
\nu(A\times B/{\mathcal A}_0)$, $P\times \nu$ a.e. for all $P\in
{\mathcal P}$. Then, since $f_{A\times B}$ is ${\mathcal
A}_0$-measurable, it depends only from $\omega_1$. Put
$\nu'(\omega_1,A)=f_{A\times \Omega_1}(\omega_1)$, then for $B\in
{\mathcal A}_1$ and $P\in {\mathcal P}$,
$$
\int_B\nu'(\omega_1,A)\nu(P)(d\omega_1)=\int_{\Omega \times
B}f_{A\times \Omega_1}d(P\times \nu)=P\times \nu(A\times
B)=\int_A\nu(\omega,B)P(d\omega),
$$
so that $\nu'=\nu_P'$,  $\nu(P)$ - a.e.
\end{proof}

\begin{theorem}\label{th:sufker} Let ${\mathcal P} \subset
M_1^+(\Omega, {\mathcal A})$ be dominated and let $P_0\in C(\mathcal P)$ be
such that $\mathcal P\sim P_0$.
Let $\nu :\Omega \times {\mathcal A}_1 \to [0,1]$ be a Markov kernel. Then
$\nu$ is sufficient for $\mathcal P$ if and only if
$\nu$ is sufficient for $\{P,P_0\}$ for every $P\in\mathcal P$.
\end{theorem}

\begin{proof}  Let us denote $$\mathcal P\times \nu:=
\{P\times \nu: P\in\mathcal P\}.$$
By Lemma \ref{le:sufalg},
$\nu$ is sufficient for all $\{P, P_0\}$ if and
only if the sub-$\sigma$-algebra ${\mathcal A}_0$ is sufficient
for all $\{ P\times \nu ,P_0\times \nu\}$. Clearly,
$P_0\times \nu\in C(\mathcal P\times \nu)$ and if we have
$\mathcal P\times \nu\sim P_0\times \nu$, then, by Theorem \ref{th:sufalg},
${\mathcal A}_0$ is sufficient for $\mathcal P\times \nu$ and
therefore $\nu$ is sufficient for ${\mathcal P}$. It is enough to
prove that $\{ P\times \nu :P\in {\mathcal P}\}$ is dominated by
$P_0\times \nu$.

For this, fix $\epsilon >0$ and let $A\in {\mathcal A}$,
$B\in {\mathcal A}_1$. By Kolmogorov inequality,
\begin{equation}\label{eq:kolmog}
P_0\times \nu(A\times B)=\int_A\nu(\omega,B)P_0(d\omega)\geq kP_0(A\cap \{
\nu(\omega,B)\geq k\})
\end{equation}
for all $k\geq 0$, moreover, since $\nu(\omega,B)\leq 1$,
\begin{eqnarray*}
P\times \nu(A\times B)&=&\int_A\nu(\omega,B)P(d\omega)=\int_{A\cap
\{ \nu(\omega B)\geq k\}}\nu(\omega,B)P(d\omega)\\
&+& \int_{A\cap \{ \nu(\omega,B)<k\}}\nu(\omega,B)P(d\omega)\leq
P(A\cap \{ \nu(\omega,B)\geq k\})+k.
\end{eqnarray*}

Since $P \ll P_0$, there is some $\delta >0$, such that
$P(A)<\epsilon/2$ if $P_0(A)<2\delta/\epsilon$. Put $k=\epsilon/2$ in
(\ref{eq:kolmog}), then $P_0\times \nu(A\times B)<\delta$ implies that
$P_0\cap \{ \nu(\omega,B)\geq \epsilon/2\})<2\delta/\epsilon$ and $P\times
\nu(A\times B)<\epsilon$.

Let now $C\in {\mathcal A}\otimes {\mathcal A}_1$ be such that
$P_0\times \nu(C)<\delta/2$. Then, since $C$ can be approximated by
rectangles, there are some $A\in {\mathcal A}$ and $B\in{\mathcal
A}_1$, such that $A\times B\supset C$ and $P_0\times \nu(A\times
B)<\delta$. This implies that $P(C)\leq P\times \nu(A\times
B)<\epsilon$.
\end{proof}

A comparison of Blackwell sufficiency and sufficiency is given in
the following theorem (\cite[Theorem 22.11]{Hey}).

\begin{theorem}\label{th:compar} Let $(\Omega, {\mathcal A},
{\mathcal P})$ be an experiment, $(\Omega_1, {\mathcal A}_1)$ a
measurable space and $\nu :\Omega\times {\mathcal A}_1\to [0,1]$ a
Markov kernel.
\begin{enumerate}
\item[{\rm (i)}] If $(\Omega, {\mathcal A}, {\mathcal P})$ is $\mu$
-dominated by a $\sigma$-finite measure $\mu$ on $(\Omega, {\mathcal
A})$ and $\nu$ is Blackwell sufficient, then $\nu$ is sufficient.
\item[{\rm(ii)}] If $(\Omega_1,{\mathcal A}_1, \nu({\mathcal P}))$ is
$\mu_1$-dominated by a $\sigma$-finite measure $\mu_1$,
$(\Omega,{\mathcal A})$ is a standard Borel space and $\nu$ is
sufficient for ${\mathcal P}$, then $\nu$ is also Blackwell
sufficient.
\end{enumerate}
\end{theorem}

\begin{proof} (i) can be proved  from Theorems \ref{th:mon},
\ref{th:klc} and \ref{th:sufker},
(ii) follows from Remark \ref{re:suff} and Theorem \ref{th:stand}.
\end{proof}

We can list  the results of the present section as follows.

\begin{corollary}\label{co:2} Let ${\mathcal P}\subset
M_1^+(\Omega,{\mathcal A}_1)$ be a dominated family and let $P_0\in
C({\mathcal P})$ be such that ${\mathcal P}\sim P_0$. Let $\nu
:\Omega \times {\mathcal A}_1 \to [0,1]$ be a Markov kernel. Then
the following are equivalent.
\begin{enumerate}
\item[{\rm(i)}] $\nu$ is pairwise sufficient for ${\mathcal P}$.
\item[{\rm(ii)}] $\nu$ is sufficient for $\{ P,P_0\}$ for each $P\in
{\mathcal P}$.
\item[{\rm(iii)}] For all $P\in {\mathcal P}$ and all convex
functions $f: [0,\infty)\to {\mathbb R}$, we have
$$
D_f(P,P_0)=D_f(\nu(P), \nu(P_0)).
$$
\item[{\rm(iv)}] There is a strictly convex function
$f:[0,\infty)\to {\mathbb R}$, such that for all $P\in {\mathbb P}$
$$
D_f(P,P_0)=D_f(\nu(P),\nu(P_0)) < \infty.
$$
\item[{\rm(v)}] $\nu$ is sufficient for ${\mathcal P}$.
\item[{\rm(vi)}] There is a weak Markov kernel $\nu':\Omega_1\times
{\mathcal A}_1\to [0,1]$ with respect to $\{ \nu(P): P\in {\mathcal
P}\}$, such that $\nu'(\nu(P))=P$ for all $P\in {\mathcal P}$. If
$(\Omega, {\mathcal A})$ is a standard Borel space, then $\nu'$ is a
Markov kernel and $\nu$ is Blackwell sufficient for $\mathcal P$.
\end{enumerate}
\end{corollary}
\begin{proof} The equivalence of (i) -(v) follows directly
from our previous results, (v)$\implies$(vi) follows from the
Remark \ref{re:suff}. The implication
(vi)$\implies$ (v) follows from Theorems \ref{th:mon} and \ref{th:klc}.
\end{proof}

\subsection{Application to fuzzy quantum observables}

Recall that if $m_0$ is a faithful state on $L$ then, for every
state $m$  and every observable $\xi$ on $L$, it holds $m\circ \xi
\ll m_0\circ \xi$. Therefore $\mathcal P=\{ m\circ \xi :m\in
{\mathcal S}\}$ is a dominated family, with $\mathcal P\sim
m_0\circ\xi$. Applying Corollary \ref{co:2} and Theorem
\ref{th:compar}, we obtain the following theorem.

\begin{theorem}\label{th:final} Let $L$ be a $\sigma$-orthocomplete
effect algebra with an order determining system ${\mathcal S}$ of
$\sigma$-additive states, $\xi$ and $\eta$ be real observables on
$L$ such that $\xi \preceq \eta$ with a confidence measure $\nu$,
and let there exist a faithful state $m_0\in {\mathcal S}$. The
following conditions are equivalent.
\begin{enumerate}
\item[{\rm(i)}]$\nu$ is pairwise sufficient for $\{ m\circ \xi
:m\in {\mathcal S}\}$.
\item[{\rm(ii)}] $\nu$ is sufficient for $\{ m\circ \xi, m_0\circ \xi \}$
for all $m\in {\mathcal S}$.
\item[{\rm(iii)}] for all $m\in {\mathcal S}$ and all convex
functions $f:[0,\infty)\to {\mathbb R}$, we have
$$
D_f(m\circ \xi, m_0\circ \xi)=D_f(\nu(m\circ \xi),\nu(m_0\circ
\xi)).
$$
\item[{\rm(iv)}] There is a strictly convex function
$f:[0,\infty)\to {\mathbb R}$ such that for all $m\in {\mathcal S}$
$$
D_f(m\circ \xi, m_0\circ \xi)=D_f(\nu(m\circ \xi),\nu(m_0\circ
\xi))<\infty.
$$
\item[{\rm(v)}] $\nu$ is sufficient for $\{ m\circ \xi : m\in
{\mathcal S}\}$.
\item[{\rm(vi)}] $\nu$ is Blackwell sufficient for $\{ m\circ \xi : m\in
{\mathcal S}\}$.
\item[{\rm(vii)}] $\xi \sim \eta$.
\end{enumerate}
\end{theorem}


\begin{thebibliography}{xxxx}

\bibitem{BW} Barbieri, G., Weber, H., Measures on clans and on
MV-algebras, in: E. Pap (Ed.), Handbook of Measure Theory, vol. II,
Elsevier, Amsterdam, The Nederlands, 2002, pp. 911-945.

\bibitem{BLM} Busch, P., Lahti, P., Mittelstaedt P., {\it The
Quantum Theory of Measurement}, Lecture Notes in Physics,
Springer-Verlag, Berlin, Heidelberg, New York, London, Budapest,
1991.

\bibitem{Bu} Bugajski, S., Statistical maps I. Basic properties,
{\it Math. Slovaca} {\bf 51} (2001), 321-342.


\bibitem{BHS} Bugajski, S., Hellwig, K.E., Stulpe, W., On fuzzy
random variables and statistical maps, {\it Rep. Math. Phys.} {\bf
41} (1998), 1-11.

\bibitem{Bud} Busemi, F., D'Ariano, G.M., Keyl, M., Perinotti, P.,
Werner, R.F., Ordering of measurements according to quantum noise,
Lecture on QUIT, Budmerice, 2 December 2004

\bibitem{BK} Butnariu, D., Klement, E., Triangular-norm-based
measures and their Markov kernel representation, J. Math. anal.
Appl. {\bf 162}(1991), 111-143.

\bibitem{CDM} Cignoli, R., D'Ottaviano, I.M.L., Mundici, D., {\it
Algebraic fFundations of Many-Valued Reasoning}, Kluwer, Dordrecht,
2000.

\bibitem{Ch} Chang, C.C., Algebraic analysis of many valued logic,
{\it Trans. Amer. Math. Soc.} {\bf 88} (1958), 467-490.

\bibitem{ChK}Chovanec, F., K\^opka, F., Boolean D-posets, {\it Tatra Mt.
Math. Publ.} {\bf 10} (1997), 183-197.

\bibitem{Dav} Davies, E.B., {\it Quantum Theory of Open Systems},
Academic Press, London, 1976.

\bibitem{DDD} Ducho\v n, M., Dvure\v censkij, A., De Lucia, P.,
Moment problem for effect algebras, Moment problem for effect
algebras, {\it Inter. J. Theor. Phys.} {\bf 36} (1997), 1941-1958.

\bibitem{Dv} Dvure\v censkij, A., Loomis-Sikorski theorem for $\sigma$-complete
MV-algebras and $\ell$-groups, {\it J. Austral. Math. Soc.} Ser. A
{\bf 68} (2000), 261-277.

\bibitem{DvPu} Dvure\v censkij, A., Pulmannov\'a, S., {\it New
Trends in Quantum Structures}, Kluwer, Dordrecht, 2000.

\bibitem{DvPu1} Dvure\v censkij, A., Pulmannov\'a, S., Difference
posets, effects and quantum measurements, {\it Inter. J. Theor.
Phys.} {\bf 33} (1994), 819-850.

\bibitem{DLPY} Dvure\v censkij, A., Lahti, P., Pulmannov\'a, S.,
Ylinen, K., Notes on a coarse grainings and functions of
observables, {\it Rep. Math. Phys.} {\bf 55} (2005), 241-248.

\bibitem{FB} Foulis, D., Bennett, M.K.,
Effect algebras and unsharp quantum logics, {\it Found. Phys.} {\bf
24} (1994), 1325-1346.

\bibitem{GG} Giuntini, R., Greuling, H.,Toward a formal language for
unsharp properties, {\it Found. Phys.} {\bf 19} (1989), 931-945.

\bibitem{Gud} Gudder, S., Lattice properties of quantum effects.
{\it J. Math. Phys.} {\bf 37} (1996), 2637-2642.





\bibitem{HS} Halmos, P.R., Savage, L.J., Applications of the
Radon-Nikodym theorem to the theory of sufficient statistics, {\it
Ann. Math. Statist.} {\bf 20} (1949), 225-241.


\bibitem{He} Heinone, T., Optimal measurement in quantum mechanics,
{\it Phys. Lett. A} {\bf 346} (2005), 77-86.

\bibitem{HLY} Heinonen, T., Lahti, P., Ylinen, K., Covariant fuzzy
observables and coarse-grainings, {\it Rep. Math. Phys.} {\bf 53}
(2004), 425-441.

\bibitem{Hey} Heyer, H., {\it Theory of Statistical Experiments},
Springer, New york, Heidelberg, Berlin, 1982.

\bibitem{KCh} K\^opka, F., Chovanec, F., D-posets, {\it Math.
Slovaca} {\bf 44} (1994), 21-34.



\bibitem{LM} Lahti, P.J., M\c aczi\'nski, M.J., On the order
structure of the set of effects in quantum mechanics, {\it J. Math.
Phys.} {\bf 36} (1995), 1673-1680.

\bibitem{Lo} Lo\`{e}ve, M., {\it Probability Theory}, D. van
Nostrand Co., New York, 1955.

\bibitem{LV} Liese, L., Vajda, I., {\it Convex Statistical
Distances}, Teubner-Texte zur Mathematik, Leipzig, 1987.

\bibitem{Mu} Mundici,D., Tensor product and the Loomis-Sikorski
theorem for MV-algebras, {\it Adv. Appl. Math.} {\bf 22} (1999),
227-248.

\bibitem{PtPu} Pt\'ak, P., Pulmannov\'a, S., {\it Orthomodular
Structures as Quantum Logics}, Kluwer, Dordrecht 1991.

\bibitem{St} St\v ep\'an, J., {\it Probability Theory (Teorie pravd\v
epodobnosti, in Czech)}, Academia, Prague, 1987.

\bibitem{Str} Strasser, H., {\it Mathematical Theory  of
Statistics}, Walter de Gruyter, Berlin, 1985.

\bibitem{Var} Varadarajan, V.S., {\it Geometry of Quantum theory},
Springer-Verlag, Berlin, 1985.

\end{thebibliography}
\end{document}